\documentclass[twocolumn]{aastex631}
\usepackage{mathrsfs }
\usepackage{natbib}
\usepackage{amssymb}
\usepackage{apjfonts}
\usepackage{ulem} 
\usepackage{graphicx}
\begin{document}
\title{The AURORA Survey: A New Era of Emission-line Diagrams with {\it JWST}/NIRSpec}

\author[0000-0003-3509-4855]{Alice E. Shapley}\affiliation{Department of Physics \& Astronomy, University of California, Los Angeles, 430 Portola Plaza, Los Angeles, CA 90095, USA}
\email{aes@astro.ucla.edu}

\author[0000-0003-4792-9119]{Ryan L. Sanders}\affiliation{Department of Physics and Astronomy, University of Kentucky, 505 Rose Street, Lexington, KY 40506, USA}

\author[0000-0001-8426-1141]{Michael W. Topping}\affiliation{Steward Observatory, University of Arizona, 933 N Cherry Avenue, Tucson, AZ 85721, USA}

\author[0000-0001-9687-4973]{Naveen A. Reddy}\affiliation{Department of Physics \& Astronomy, University of California, Riverside, 900 University Avenue, Riverside, CA 92521, USA}

\author[0000-0002-4153-053X]{Danielle A. Berg}\affiliation{Department of Astronomy, The University of Texas at Austin, 2515 Speedway, Stop C1400, Austin, TX 78712, USA}

\author[0000-0002-4989-2471]{Rychard J. Bouwens}\affiliation{Leiden Observatory, Leiden University, NL-2300 RA Leiden, Netherlands}

\author[0000-0003-2680-005X]{Gabriel Brammer}\affiliation{Niels Bohr Institute, University of Copenhagen, Lyngbyvej 2, DK2100 Copenhagen \O, Denmark}

\author[0000-0002-1482-5818]{Adam C. Carnall}\affiliation{Institute for Astronomy, University of Edinburgh, Royal Observatory, Edinburgh, EH9 3HJ, UK}

\author[0000-0002-3736-476X]{Fergus Cullen}\affiliation{Institute for Astronomy, University of Edinburgh, Royal Observatory, Edinburgh, EH9 3HJ, UK}

\author[0000-0003-2842-9434]{Romeel Dav\'e}\affiliation{Institute for Astronomy, University of Edinburgh, Royal Observatory, Edinburgh, EH9 3HJ, UK}

\author{James S. Dunlop}\affiliation{Institute for Astronomy, University of Edinburgh, Royal Observatory, Edinburgh, EH9 3HJ, UK}

\author[0000-0001-7782-7071]{Richard S. Ellis}\affiliation{Department of Physics \& Astronomy, University College London. Gower St., London WC1E 6BT, UK}

\author[0000-0003-4264-3381]{N. M. F\"orster Schreiber}\affiliation{Max-Planck-Institut f\"ur extraterrestrische Physik (MPE), Giessenbachstr.1, D-85748 Garching, Germany}

\author[0000-0002-0658-1243]{Steven R. Furlanetto}\affiliation{Department of Physics \& Astronomy, University of California, Los Angeles, 430 Portola Plaza, Los Angeles, CA 90095, USA}

\author[0000-0002-3254-9044]{Karl Glazebrook}\affiliation{Centre for Astrophysics and Supercomputing, Swinburne University of Technology, P.O. Box 218, Hawthorn, VIC 3122, Australia}

\author[0000-0002-8096-2837]{Garth D. Illingworth}\affiliation{Department of Astronomy and Astrophysics, UCO/Lick Observatory, University of California, Santa Cruz, CA 95064, USA}

\author[0000-0001-5860-3419]{Tucker Jones}\affiliation{Department of Physics and Astronomy, University of California Davis, 1 Shields Avenue, Davis, CA 95616, USA}

\author[0000-0002-7613-9872]{Mariska Kriek}\affiliation{Leiden Observatory, Leiden University, NL-2300 RA Leiden, Netherlands}

\author[0000-0003-4368-3326]{Derek J. McLeod}\affiliation{Institute for Astronomy, University of Edinburgh, Royal Observatory, Edinburgh, EH9 3HJ, UK}

\author{Ross J. McLure}\affiliation{Institute for Astronomy, University of Edinburgh, Royal Observatory, Edinburgh, EH9 3HJ, UK}

\author[0000-0002-7064-4309]{Desika Narayanan}\affiliation{Department of Astronomy, University of Florida, 211 Bryant Space Sciences Center, Gainesville, FL, USA}

\author[0000-0001-5851-6649]{Pascal Oesch}\affiliation{Department of Astronomy, University of Geneva, Chemin Pegasi 51, 1290 Versoix, Switzerland}

\author[0000-0003-4464-4505]{Anthony J. Pahl}\affiliation{The Observatories of the Carnegie Institution for Science, 813 Santa Barbara Street, Pasadena, CA 91101, USA}

\author{Max Pettini}\affiliation{Institute of Astronomy, Madingley Road, Cambridge CB3 OHA, UK}

\author[0000-0001-7144-7182]{Daniel Schaerer}\affiliation{Department of Astronomy, University of Geneva, Chemin Pegasi 51, 1290 Versoix, Switzerland}

\author{Daniel P. Stark}\affiliation{Steward Observatory, University of Arizona, 933 N Cherry Avenue, Tucson, AZ 85721, USA}

\author[0000-0002-4834-7260]{Charles C. Steidel}\affiliation{Cahill Center for Astronomy and Astrophysics, California Institute of Technology, MS 249-17, Pasadena, CA 91125, USA}

\author[0000-0001-5940-338X]{Mengtao Tang}\affiliation{Steward Observatory, University of Arizona, 933 N Cherry Avenue, Tucson, AZ 85721, USA}

\author[0000-0003-1249-6392]{Leonardo Clarke}\affiliation{Department of Physics \& Astronomy, University of California, Los Angeles, 430 Portola Plaza, Los Angeles, CA 90095, USA}

\author[0000-0002-7622-0208]{Callum T. Donnan}\affiliation{Institute for Astronomy, University of Edinburgh, Royal Observatory, Edinburgh, EH9 3HJ, UK}

\author{Emily Kehoe}\affiliation{Department of Physics \& Astronomy, University of California, Los Angeles, 430 Portola Plaza, Los Angeles, CA 90095, USA}

\shortauthors{Shapley et al.}

\shorttitle{A New Era of Emission-line Diagrams}

\begin{abstract}
We present results on the emission-line properties of $z=1.4-7.5$ star-forming galaxies
in the Assembly of Ultradeep Rest-optical Observations Revealing Astrophysics (AURORA) Cycle~1 {\it JWST}/NIRSpec program.
Based on its depth, continuous wavelength coverage from 1--5$\mu$m, and medium spectral resolution ($R\sim 1000$), AURORA includes
detections of a large suite of nebular emission lines spanning a broad range in rest-frame wavelength. We investigate
the locations of AURORA galaxies in multiple different emission-line diagrams, including traditional ``BPT" diagrams of
[OIII]$\lambda 5007$/H$\beta$ vs. [NII]$\lambda 6583$/H$\alpha$, [SII]$\lambda\lambda6717,6731$/H$\alpha$, and [OI]$\lambda 6300$/H$\alpha$,
and the ``ionization-metallicity" diagram of [OIII]$\lambda 5007$/[OII]$\lambda3727$ ($O_{32}$) vs. ([OIII]$\lambda 5007$+[OII]$\lambda3727$)/H$\beta$ ($R_{23}$). We also consider a bluer rest-frame ``ionization-metallicity" diagram introduced recently to 
characterize $z>10$ galaxies: [NeIII]$\lambda 3869$/[OII]$\lambda3727$ vs. ([NeIII]$\lambda 3869$+[OII]$\lambda3727$)/H$\delta$;
as well as longer-wavelength diagnostic diagrams extending into the rest-frame near-IR: [OIII]$\lambda 5007$/H$\beta$ vs.
[SIII]$\lambda\lambda 9069,9532$/[SII]$\lambda\lambda6717,6731$ ($S_{32}$); and He~I$\lambda 1.083\mu$m/Pa$\gamma$ 
and [SIII]$\lambda 9532$/Pa$\gamma$ vs. [FeII]$\lambda 1.257\mu$m/Pa$\beta$. With a significant boost in signal-to-noise
and large, representative samples of individual galaxy detections, the AURORA emission-line diagrams presented here definitively
confirm a physical picture in which chemically-young, $\alpha$-enhanced, massive stars photoionize the ISM in distant galaxies
with a harder ionizing spectrum at fixed nebular metallicity than in their $z\sim0$ counterparts. We also
uncover previously unseen evolution prior to $z\sim 2$ in the [OIII]$\lambda 5007$/H$\beta$ vs. [NII]$\lambda 6583$/H$\alpha$
diagram, which motivates deep NIRSpec observations at even higher redshift. Finally, we present the first statistical
sample of rest-frame near-IR emission-line diagnostics in star-forming galaxies at high redshift. In order to truly interpret
rest-frame near-IR line ratios including [FeII]$\lambda 1.257\mu$m, we must obtain better constraints on dust depletion
in the high-redshift ISM.

\end{abstract}

\section{Introduction}

Recombination and collisionally-excited emission lines from nearby star-forming regions and galaxies provide the key to understanding the interplay between massive stars, gas, dust, and heavy elements in the ionized interstellar medium. In the Orion Nebula, for example, several hundred optical emission lines have been measured \citep{esteban2004}, enabling an exquisitely detailed view of the density and temperature structure in the gas, as well as the mixture of chemical elements. Similar analyses can be extended to extragalactic H~II regions and low-redshift star-forming galaxies \citep[e.g.,][]{izotov2006,berg2015,mendezdelgado2023}.

With the advent of near-IR spectrographs on 8-10 meter telescopes on the ground and the {\it HST}/WFC/IR grism in space, it was possible to extend rest-optical nebular emission-line measurements to samples of star-forming galaxies at $z\sim 1-3$ \citep[e.g.,][]{steidel2014,shapley2015,sanders2016,momcheva2016,kashino2017,forsterschreiber2019,sanders2021}. In such studies, the set of detected emission lines  has typically been limited to the brightest rest-optical lines, including [OII]$\lambda3727$, H$\beta$, [OIII]$\lambda\lambda5007,4959$, H$\alpha$, [NII]$\lambda6583$, and [SII]$\lambda\lambda6717,6731$ \citep[but see, e.g.,][]{sanders2023a,clarke2023,jeong2020,backhaus2022}. Furthermore, the strongly wavelength-dependent and bright sky background due to the forest of OH vibration-rotation transitions at $1-2.5 \mu{\rm m}$ leads to significant challenges in obtaining robust measurements from the ground for all but the very brightest of the rest-optical emission lines (e.g., H$\alpha$, [OIII]$\lambda5007$). 

Despite these observational hurdles, significant progress was achieved over the past decade based on multi-object near-IR spectroscopic surveys out to $z\sim 3$.  Relations among metallicity, mass, star-formation rate (SFR), dynamical properties, and dust attenuation have been constructed for star-forming galaxies, demonstrating
the evolution at fixed stellar mass towards higher SFR, lower metallicity, and lower ratio of rotational-to-random dynamical support, with increasing redshift \citep{shapley2015,shivaei2015,kashino2017,wisnioski2019,forsterschreiber2020,shapley2022}. Furthermore, galaxies seem to follow the same manifold of stellar mass, SFR, and metallicity at $z=0-3$, suggesting the same equilibrium is in place among inflows, outflows, and star formation \citep{sanders2021}. Finally, an essential characteristic of distant star-forming galaxies emerging from these spectroscopic studies is their chemical youth, encapsulated by evidence for non-solar, $\alpha$-enhanced chemical abundance patterns. This evidence comes in many forms, including a comparison of nebular oxygen and stellar iron abundances \citep[e.g.,][]{steidel2016,topping2020,cullen2021,stanton2024}; the loci of high-redshift galaxies in multiple rest-optical emission-line diagnostic diagrams \citep[e.g.][]{steidel2014,shapley2019,sanders2020a,jeong2020,clarke2023}; and photoionization modeling of strong nebular emission-line ratios incorporating direct metallicity estimates \citep[e.g.][]{sanders2020b}.

The infrared spectroscopic capabilities of {\it JWST} represent a transformational advance for the study of ionized gas in star-forming galaxies at high redshift. The redshift frontier for rest-optical emission-line measurements has been extended from $z\sim 3$ out to $z\sim 12$ \citep[e.g.,][]{shapley2023a,sanders2023b,sanders2024a,cameron2023,nakajima2023,robertsborsani2024,castellano2024}. Furthermore, the unprecedented wavelength baseline for high signal-to-noise (S/N) infrared spectroscopy
has highlighted the promise of extending diagnostic nebular studies into the rest-frame
near-IR for galaxies observed during the epoch of peak star formation in the universe ($z\sim 2-3$) \citep{brinchmann2023,calabro2023}. At the other end of the rest-frame optical, blue rest-optical emission-line diagnostics \citep[e.g.,][]{bunker2023} have
been obtained for the most extreme galaxies at $z>10$. While not the focus of the current work, {\it JWST} has also enabled unprecedented rest-UV spectroscopic diagnostics of galaxies at $z\sim 4-12$ \citep[e.g.,][]{tang2023,williams2023,castellano2024,topping2024}. Thus far, these new emission-line diagnostic diagrams are populated by extremely small high-redshift samples, with typically modest S/N. Alternatively, data from stacked measurements is plotted, which increases the S/N of faint features at the cost of obscuring the variation within a given dataset \citep[e.g,][]{sanders2023b}.

The Assembly of Ultradeep Observations Revealing Astrophysics (AURORA) Cycle~1 {\it JWST}/NIRSpec program was designed to detect faint auroral lines from
ionized oxygen, sulfur, and nitrogen for $z>1.4$ galaxies, which can be used to infer electron temperatures
and, in turn, direct chemical abundances. As such, AURORA consists of  long exposures with
{\it JWST}/NIRSpec using each of the three medium-resolution ($R\sim1000$) gratings. Because of their superior depth, the AURORA data yield significantly higher S/N measurements than in previous work, including emission-line excitation sequences for individual high-redshift galaxies spanning multiple diagnostic emission-line ratio diagrams (e.g., 
[OIII]$\lambda5007$/H$\beta$ vs. [NII]$\lambda6583$/H$\alpha$). 
Along with deep integration times,
AURORA was also designed to obtain continuous wavelength coverage from $1-5 \mu{\rm m}$. The AURORA
dataset thus yields both rest-frame blue-optical emission-line diagnostics highlighted for interpreting unique $z>10$ spectra, and also the first statistical sample of rest-frame near-IR diagnostics for star-forming galaxies at $z>1$. We demonstrate how these diagnostics provide novel insights into the chemical abundance patterns and interplay between massive stars and gas in distant star-forming galaxies.

In \S\ref{sec:obs}, we describe
the AURORA observations, sample, data reduction, flux calibration including slit-loss corrections, and measurements. In \S\ref{sec:results}, we present results on familiar emission-line diagnostic diagrams, extended towards higher redshift, and novel emission-line diagrams spanning a broader wavelength baseline and measured for the first time with actual statistical power. 
In \S\ref{sec:discussion}, we consider the implications of these new measurements and discuss
future directions.
Throughout, we adopt cosmological parameters of
$H_0=70\mbox{ km  s}^{-1}\mbox{ Mpc}^{-1}$, $\Omega_{\rm m}=0.30$, and
$\Omega_{\Lambda}=0.7$, and a \citet{chabrier2003} IMF.

\section{Observations}
\label{sec:obs}
\subsection{AURORA NIRSpec Observations}
\label{sec:obs-nirspec}
We use NIRSpec Micro Shutter Assembly (MSA) data from the
AURORA program (Program ID: 1914; Co-PIs Shapley and Sanders). The AURORA NIRSpec observations
we analyzed consisted of 2 pointings, one in the COSMOS field
(PA=71.82 degrees) and
one in GOODS-N (PA=70.34 degrees). Both pointings utilized the grating/filter
combination of G140M/F100LP, G235M/F170LP, and G395M/F290LP, which
provide a spectral resolution of $R\sim 1000$
over the wavelength range approximately $1-5\mu$m.
In each pointing, the G140M/F100LP combination was observed for 44204 seconds
(30 exposures of 20 groups and the NRSIRS2 readout mode),
the G235M/F170LP combination for 28886 seconds (30 exposures of 65 groups and the
NRSIRS2RAPID readout mode), and the G395M/F290LP combination
for 15056 seconds (6 exposures of 85 groups and the NRSIRS2RAPID readout mode). The readout modes for each grating/filter were chosen to conform to the data volume allocation.
Based on the NIRSpec Exposure Time Calculator, the exposure time for each grating was designed to yield
uniform $\sim5\sigma$ emission-line sensitivity of $10^{-18}\mbox{ erg}\mbox{ s}^{-1}\mbox{ cm}^{-2}$.
A 3-point nod pattern was adopted for
each observation, and each MSA ``slit" consisted of 3 microshutters.

\subsection{AURORA Target Sample}
\label{sec:obs-targets}
The COSMOS and GOODS-N masks included, respectively, 46 and 51 galaxy targets.  Once the PA for each MSA pointing
was assigned to our program, we used the NIRSpec MSA Planning Tool (MPT)
to optimize the number of targets
predicted to yield a significant detection for at least one auroral
oxygen emission line (i.e., [OIII]$\lambda4363$ and/or [OII]$\lambda\lambda 7320,7330$).
In the considerable time that elapsed between the submission of our Cycle 1 proposal in November 2020
and its execution in late 2023/early 2024, updated empirical information became available regarding the
sensitivity of NIRSpec. Scaling from NIRSpec $R\sim1000$ measurements from the CEERS Early Release Science (ERS)
program \citep{finkelstein2023}, we
therefore lowered the threshold for ``detection" to
$5\times 10^{-19}\mbox{ erg s}^{-1}\mbox{cm}^{-2}$ ($3\sigma$).
In identifying potential auroral-line targets,
we utilized publicly-available multi-wavelength photomety and
redshifts from the 3D-HST survey \citep{skelton2014},
rest-optical emission-line catalogs of the MOSFIRE Deep Evolution Field (MOSDEF) survey \citep{kriek2015},
and the Extreme Emission-line Galaxy (EELG) catalogs of \citet{tang2019}.
We used the set of measured strong rest-frame optical nebular emission lines in the MOSDEF
and 3D-HST EELG catalogs, the relationships between strong-line ratios
and [OIII] electron temperatures ($T_3$ and $T_2$ for the ionization zones
traced, respectively, by [OIII]$\lambda5007$ and [OII]$\lambda3727$)
derived in \citet{sanders2017} and \citet{sanders2021}, and the relationships among the intensities of several
auroral lines measured in local H~II regions \citep[e.g.,][]{berg2020} to predict
$T_3$ and $T_2$, and, accordingly the intensities of the following auroral
features: [OIII]$\lambda 4363$, [SIII]$\lambda 6312$,
[OII]$\lambda\lambda 7320,7330$, [NII]$\lambda 5755$, and [SII]$\lambda 4068,4076$.
To these samples, we added galaxies with pre-{\it JWST} spectroscopic measurements of [OIII]$\lambda5007$
flux greater than or equal to $10^{-16}\mbox{ erg s}^{-1}\mbox{cm}^{-2}$, or with SEDs
indicating a photometric excess at the wavelength of  [OIII]$\lambda5007$ corresponding to
a line flux greater than or equal to $10^{-16}\mbox{ erg s}^{-1}\mbox{cm}^{-2}$.
For objects with $f(\mbox{[OIII]}\lambda5007)\geq 10^{-16}\mbox{ erg s}^{-1}\mbox{cm}^{-2}$,
we predicted a detection of [OIII]$\lambda4363$ at $\geq3\sigma$ if
$T_3\geq10,000$K, based on our limiting line sensitivity.

These criteria resulted in 249 (210) potential auroral-line targets in COSMOS (GOODS-N), ranging
in redshift from $1.2< z \leq 4.65$. These potential targets were prioritized using
the following scheme (wherein we note that galaxies at $z\geq 1.63$ receive higher priority given
that [OII]$\lambda3727$ is covered by NIRSpec in this redshift range):

\begin{enumerate}
\item $z \geq 1.63$ and both [OIII]$\lambda
$4363 and [OII]$\lambda\lambda7320,7330$ have estimated flux  $\geq7\times10^{-19}\mbox{ erg s}^{-1}\mbox{cm}^{-2}$

\item  $z \geq 1.63$ and (both [OIII]$\lambda
$4363 and [OII]$\lambda\lambda7320,7330$ have estimated flux between $5\times10^{-19}\mbox{ erg s}^{-1}\mbox{cm}^{-2}$ and $7\times10^{-19}\mbox{ erg s}^{-1}\mbox{cm}^{-2}$) or (only one of [OIII]$\lambda
$4363 or [OII]$\lambda\lambda7320,7330$ have estimated flux $\geq7\times10^{-19}\mbox{ erg s}^{-1}\mbox{cm}^{-2}$); or $z < 1.63$ and both [OIII]$\lambda
$4363 and [OII]$\lambda\lambda7320,7330$ have estimated flux $\geq7\times10^{-19}\mbox{ erg s}^{-1}\mbox{cm}^{-2}$

\item $z\geq 1.63$ and [OIII]$\lambda5007$ flux $\geq 1\times10^{-16}\mbox{ erg s}^{-1}\mbox{cm}^{-2}$ from spectroscopy or estimated to have [OIII]$\lambda5007$ flux $\geq 1\times10^{-16}\mbox{ erg s}^{-1}\mbox{cm}^{-2}$ from photometric excess; $z < 1.63$ and (both [OIII]$\lambda4363$ and [OII]$\lambda\lambda7320,7330$ have estimated flux between $5\times10^{-19}\mbox{ erg s}^{-1}\mbox{cm}^{-2}$ and $7\times10^{-19}\mbox{ erg s}^{-1}\mbox{cm}^{-2}$) or (only one of [OIII]$\lambda
$4363 or [OII]$\lambda\lambda7320,7330$ have estimated flux $\geq7\times10^{-19}\mbox{ erg s}^{-1}\mbox{cm}^{-2}$)

\item $z\geq1.63$ and only one of [OIII]$\lambda
$4363 or [OII]$\lambda\lambda7320,7330$ have estimated flux between $5\times10^{-19}\mbox{ erg s}^{-1}\mbox{cm}^{-2}$ and $7\times10^{-19}\mbox{ erg s}^{-1}\mbox{cm}^{-2}$; $z < 1.63$ and [OIII]$\lambda5007$ flux $\geq 1\times10^{-16}\mbox{ erg s}^{-1}\mbox{cm}^{-2}$ from spectroscopy or estimated to have [OIII]$\lambda5007$ flux $\geq1\times10^{-16}\mbox{ erg s}^{-1}\mbox{cm}^{-2}$ from photometric excess

\end{enumerate}

Using the MPT, we were able to assign slits to 16 (20) auroral-line targets on the COSMOS (GOODS-N) mask, and optimized the mask design such that the lines of interest (auroral features and their strong nebular counterparts) fell on the NIRSpec detectors.
These 36 targets span a redshift range of $1.39 \leq z \leq 4.41$, with a median of $z_{\rm med}=2.33$;
all but 3 are at $z\geq 1.6$ and therefore have coverage of [OII]$\lambda3727$.

Based on the science goals of the AURORA program, we included additional sources in each of the COSMOS
and GOODS-N AURORA masks. In COSMOS, we made use of the photometric redshift catalog from the PRIMER
survey (J. Dunlop et al. 2024, in prep.), which is defined down to a F356W limit of $m_{\rm AB,356W}=28$. We added 7 sources at $z_{\rm phot}>6$, 1 quiescent galaxy candidate at $z\sim 2$, and 22 galaxies at $1.5 \leq z_{\rm phot} \leq 5$
and $m_{\rm AB,356W}\leq 27$. In GOODS-N, we filled the mask with photometric candidates and spectroscopically-confirmed galaxies from the FRESCO survey (two H$\alpha$ emitters at $5\leq z\leq 6$, one [OIII] emitter at $7.2$; and 3 quiescent galaxy candidates at $2\leq z< 3$); 4 $z>6$ photometric candidates from the JADES survey \citep{hainline2023} and 6 the literature \citep{jung2020,finkelstein2015,bouwens2015}; and 15 galaxies at $1.5 \leq z_{\rm phot} \leq 5$ from the publicly available GOODS-N photometric redshift catalog in the Dawn JWST Archive\footnote{https://dawn-cph.github.io/dja/index.html} \citep{heintz2024}. In total we assigned slits to 97 AURORA targets.

\subsection{Data Reduction}
\label{sec:obs-redux}
We obtained final two-dimensional (2D) spectra using a data reduction pipeline composed of standard STScI tools in addition to custom routines developed for the AURORA survey.
Each raw uncalibrated exposure was processed using the \texttt{calwebb\_detector1} routine within version 1.13.4 of the \texttt{jwst} pipeline \citep{jwstpipeline}.
This step first removed the detector bias and dark current signal, and corrected for detector gain.
We then applied a linearity correction to the images, and derived a count rate for each pixel from the non-destructive readout groups.
In calculating the count rates, we mitigated significant jumps in subsequent readout groups, such as those resulting from cosmic rays and including large events identified as `snowballs.'
Next, we corrected each of the resulting count rate files for $1/f$ noise using the \texttt{nsclean} package \citep{rauscher2024}.
We extracted the 2D spectrum traced out from each slit from the corrected rate files, and applied a flat-field correction, wavelength solution, and absolute photometric calibration to each spectrum using reference files from the updated Calibration Reference Data System (CRDS) context (\texttt{jwst\_1193.pmap}) based on in-flight estimates.
As each grating, detector, and nod position was observed twice with the same pointing center and MSA configuration, we compared the repeated exposures to identify any transient artifacts (e.g., remaining cosmic rays or hot pixels) not excluded by earlier stages of the reduction.
The identified pixels were flagged as to not be utilized for the final spectrum.
Finally, the cutout spectrum from each slit was rectified, and interpolated on to a common wavelength grid before being combined following the three-dither nodding pattern described above, yielding the final 2D spectrum for each object.

One-dimensional (1D) science and error spectra were optimally extracted from the rectified 2D spectra. During extraction,
the spatial profile was based on the brightest emission line in each grating (in 88\% of grating spectra), or the integrated continuum profile if no emission lines were detected (in 10.5\% of grating spectra). Otherwise, in a very small minority of cases ($\sim1$\%) a blind extraction was performed based on the spatial profiles in the other grating(s).  Any artifacts or cosmic rays in the extracted trace not removed during the 2D reduction were manually flagged during the 1D extraction and masked during spectral fitting.

\subsection{Slit Loss Correction}
\label{sec:obs-slitloss}

The combination of a small NIRSpec MSA microshutter width ($0\farcs20$), potentially off-center location of the target within the microshutter, and a wavelength-dependent PSF necessitates accurate corrections for light falling outside the microshutter and 1D extraction window.  These ``slit losses'' were computed as follows (see \citealt{reddy2023a} for further details).  Using the NIRCam F115W imaging, a $12\arcsec \times 12\arcsec$ subimage centered on the target was extracted.  The corresponding segmentation map was used to mask all pixels belonging to unrelated objects within the subimage.  The subimage was then rotated based on the position angle of the NIRSpec MSA observations.  This rotated subimage was then convolved with Gaussian kernels to produce the expected light profiles at longer wavelengths.  The full widths at half maximum (FWHMs) of these kernels were computed by subtracting in quadrature the FWHM of the JWST PSF at $1.15\mu$m from the FHWMs of the JWST PSFs at $\lambda = 1.2 - 5.3$\,$\mu$m, in increments of $\Delta\lambda = 0.1$\,$\mu$m.  

For 4 and 2 objects in COSMOS and GOODS-N, respectively, without NIRCam F115W imaging, the HST/WFC3 F160W image was used to generate a S\'ersic fit describing the intrinsic shape of the galaxy.  The parameters of the S\'ersic fit were obtained from \citet{vanderwel2014}, and a model was created only if the fit was deemed acceptable based on the criteria given in that study.  This model was then convolved with the JWST PSFs at $\lambda = 0.6 - 5.3$\,$\mu$m, in increments of $\Delta\lambda = 0.1$\,$\mu$m, as determined from the JWST \texttt{WebbPSF} software\footnote{\url{https://www.stsci.edu/jwst/science-planning/proposal-planning-toolbox/psf-simulation-tool}}.  Finally, a point source was assumed for 5 targets in GOODS-N without F115W imaging or an acceptable S\'ersic model fit to their F160W light profiles.  

The convolved images of the target at each wavelength were then shifted according to the expected position of the target within the microshutter and then masked.  The masking accounts for the $0\farcs20$ microshutter width, the $0\farcs07$ gap between adjacent microshutters along the cross-dispersion axis, and the center and size of the window used to extract the 2D spectrograms to 1D.  The 1D spectra were then divided by the fraction of light passing through the slit and extraction apertures as a function of wavelength.

\subsection{Flux Calibration}
\label{sec:obs-fluxcal}
The procedures for flux calibration are described in detail
in \citet{sanders2024b}. Briefly, the final flux calibration
for AURORA spectra proceeded in two stages, addressing first the relative grating-to-grating
calibration, and then the overall absolute calibration across all gratings.

The depth
of the AURORA observations results in continuum and/or emission lines detected in the overlapping
regions for G140M+G235M, G235M+G395M, or both, for a majority of AURORA targets. Overlapping
measurements were used to improve the grating-to-grating flux calibration, relative to
the output of the NIRSpec pipeline (Section~\ref{sec:obs-redux}). The G235M grating
was adopted as the reference relative to which we assessed the calibration
of the G140M and G395M spectra. The median G140M/G235M and G395M/G235M flux ratios
in overlapping wavelength regions were assessed separately for each AURORA mask,
and found to be near unity for all grating pairs except G395M/G235M in COSMOS,
which was found to be 0.80.  With a small number of exceptions, G140M or G395M spectra
were scaled by the median G140M/G235M or G395M/G235M flux ratio, respectively,
determined from the sample of line and continuum
measurements in overlapping spectral regions. Exceptions consisted of outliers,
whose grating flux ratios lay outside the $>1\sigma$ scatter of the distribution
around the median. For such objects, we forced the grating flux levels to agree
in the region of overlap. In order to assess the overall robustness of the grating-to-grating
flux calibration, we measured fluxes for 156 emission lines detected at S/N$\ge$3 in
the overlap region of neighboring gratings. For these, we found a median flux offset of
 0.1\% and an intrinsic scatter of 8\% after accounting for measurement uncertainties.

The absolute flux calibration for AURORA is obtained by scaling the spectra to match
the available multi-band photometry. The spectrum of each target in each grating was passed
through {\it JWST}/NIRCam {\it HST}/WFC3 filter transmission curves that fully overlapped
the grating of interest, to obtain synthetic photometric measurements. The ratio of
photometric flux measurements from imaging and NIRSpec spectra was calculated for
all filters in which both types of measurements had S/N$\ge$3. We applied a single scale
factor to the NIRSpec spectra to force the median of the spectrum-to-imaging flux
ratios to be unity. The median scale factor was 1.35, with a standard deviation of 0.17~dex.

\begin{figure}[t!]
\centering
\includegraphics[width=1.0\linewidth]{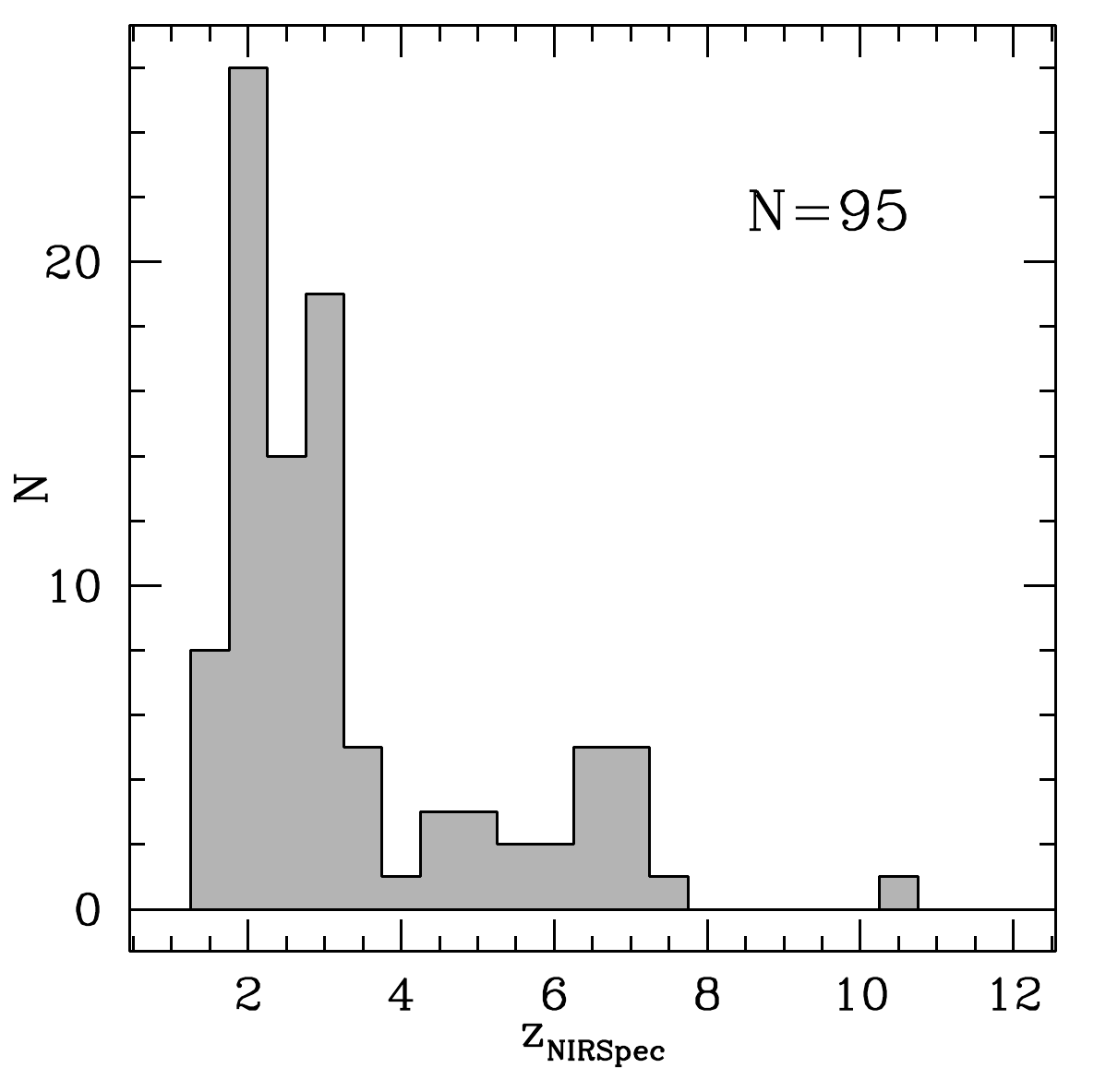}
\caption{AURORA redshift distribution. This distribution includes 36 auroral-line targets spanning a range $1.4\leq z \leq 4.5$; 18 photometric or emission-line targets at $z\geq 5$; 4 quiescent-galaxy targets at $2\leq z \leq 3$; and 37 filler targets at $1.5 \leq z \leq 5$ based on photometric redshifts. 
}
\label{fig:zhist}
\end{figure}

\begin{figure*}
\centering
\includegraphics[width=1.0\linewidth]{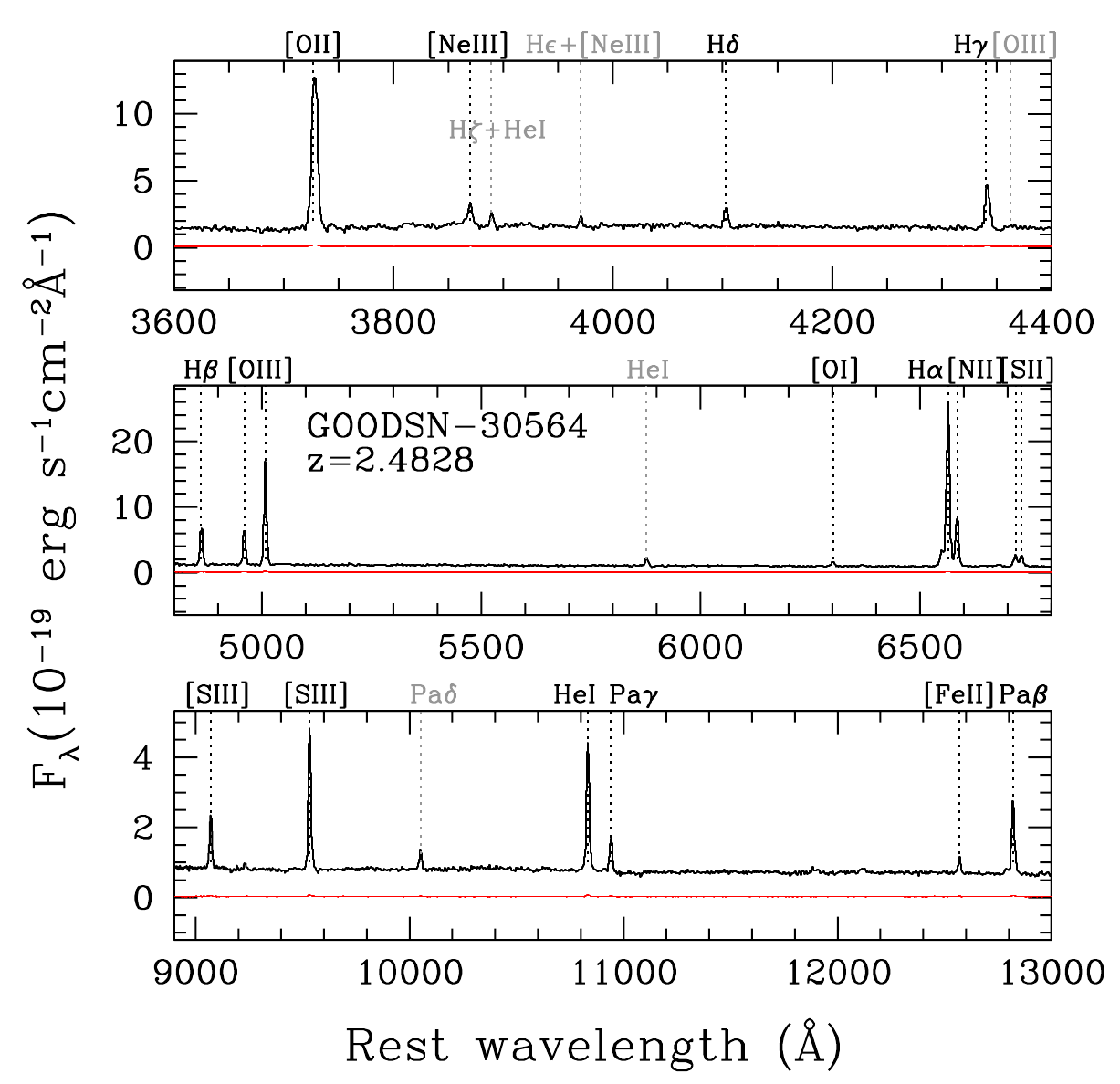}
\caption{Example AURORA spectrum, for GOODSN-30564 ($z=2.4828$). This galaxy was targeted based on the prediction that auroral emission lines ([OIII]$\lambda4363$ and [OII]$\lambda\lambda 7320, 7330$) would yield significant detections in its AURORA spectrum, and both features are successfully detected. Flux density is plotted in black, while the error spectrum is plotted in red. Black labels indicate emission lines analyzed in Section~\ref{sec:results}, while grey labels correspond to additional visible emission lines whose analysis we save for future work. The spectral windows plotted here are zoomed in to highlight features discussed in this work, but represent only a subset of the rest wavelengths covered for GOODSN-30564 by our NIRSpec set up.
}
\label{fig:spec-30564}
\end{figure*}

\begin{figure}[t!]
\centering
\includegraphics[width=1.0\linewidth]{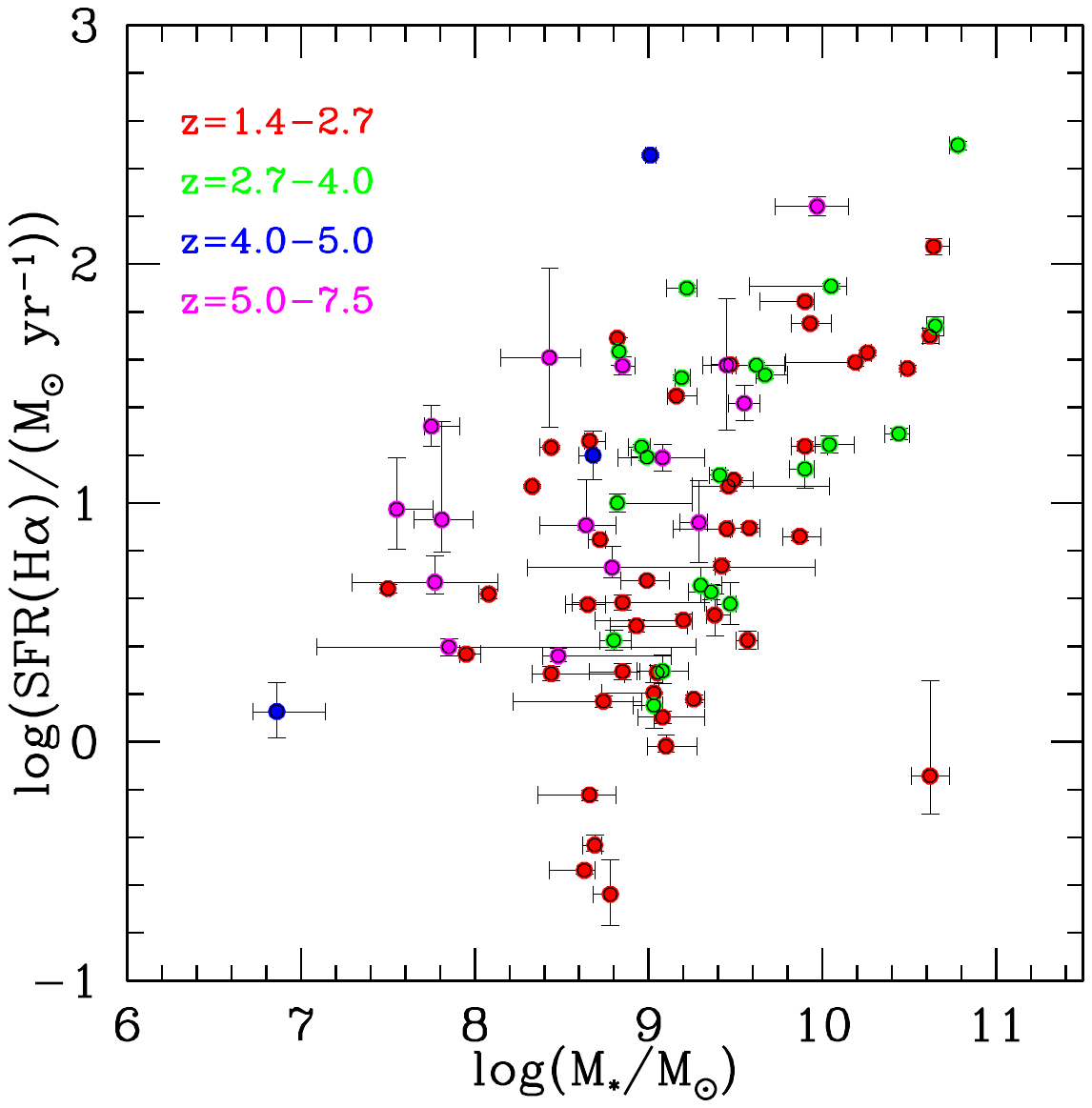}
\caption{SFR(H$\alpha$) vs. $M_*$. Red, green, blue, and magenta symbols are used, respectively, for the $1.4 \leq z < 2.7$, $2.7\leq z < 4.0$, $4.0 \leq z < 5.0$, and $z\geq 5.0$ samples. SFR(H$\alpha$) was estimated from the inferred dust-corrected H$\alpha$ luminosity, using the conversion factor from \citet{hao2011}. Galaxies flagged as AGNs or quiescent are not included here. 
}
\label{fig:ms-ha}
\end{figure}

\subsection{Emission Line Measurements}
\label{sec:obs-emline}
We measured redshifts and emission-line fluxes for 95 out of 97 galaxies targeted in the AURORA sample. The distribution of redshifts measured from NIRSpec spectra is shown in Figure~\ref{fig:zhist}.
We used single Gaussian fits for widely-separated lines; adjacent lines such as [NII]$\lambda$6548, H$\alpha$, and [NII]$\lambda$6583 were fit simultaneously with multiple Gaussians; and closely spaced lines that are blended and unresolved at $R\sim1000$ are fit with a single Gaussian. As a first pass, the underlying continuum was taken to be the best-fit SED model to the raw multi-wavelength photometry.  We then used these initial emission-line fluxes to correct the photometry for the contributions from nebular emission, and repeated the SED fitting. We then repeated the process of emission-line measurements as in the first pass, but using
the updated SED fit as the estimate of the continuum. This same best-fit SED model is used for stellar mass estimates, as described in the next section. The second pass of spectral fitting represents our final set of emission-line measurements, and (as in the first pass) automatically accounts for any underlying stellar absorption affecting nebular Balmer emission lines. Figure~\ref{fig:spec-30564} shows the final flux-calibrated spectrum of one of the auroral-line targets from the AURORA sample, GOODSN-30564 ($z=2.4828$), which boasts the detection of more than 50 individual emission lines (only a subset of which are shown within the wavelength ranges of Figure~\ref{fig:spec-30564}).

\subsection{SED Fitting}
\label{sec:obs-sed}
The spectral energy distribution (SED) of each galaxy was fit using the a large suite of publicly available photometry from JWST/NIRCam, and HST/WFC3 and ACS, with photometric catalogs downloaded from the DAWN JWST Archive. The COSMOS catalog includes the following photometric bands: (from {\it HST}/ACS and WFC3) F435W, F606W, F814W, F850LP, F105W, F125W, F140W, F160W, (and from {\it JWST}/NIRCam) F090W, F115W, F150W, F200W, F277W, F356W, F410M, and F444W. The GOODS-N catalog includes the following bands: (from {\it HST}/ACS and WFC3) F435W, F606W, F775W, F814W, F850LP, F105W, F125W, F140W, F160W, (and from {\it JWST}/NIRCam) F090W, F115W, F150W, F182M, F200W, F210M, F277W, F335M,  F356W, F410M, and F444W. In COSMOS, the footprint of our mask almost entirely overlaps the PRIMER imaging area, but 4 sources do not have PRIMER coverage. For such sources, we used photometry from the 3DHST catalog, including ground-based imaging, HST/WFC3 and ACS, and Spitzer/IRAC \citep{skelton2014}. Furthermore, for one source in COSMOS, the photometry in the DJA catalog is affected by bad pixels at wavelengths longer than 2$\mu$m. Accordingly, we used measurements from the catalog described in Donnan et al. (2024) (which are consistent with the DJA catalog at wavelengths that overlap), including the {\it HST}+{\it JWST} bands F435W, F606W, F814W, F090W, F115W, F150W, F200W,  F277W, F356W, F410M, and F444W.

For SED modeling, we used the program FAST \citep{kriek2009}, assuming
the flexible stellar population synthesis (FSPS) models of \citet{conroy2009} and a
\citet{chabrier2003} IMF, and fixed the galaxy redshift to the value determined from AURORA NIRSpec spectra. Following the methodology of \citet{reddy2018a}, we utilized two combinations
of metallicity and extinction/attenuation curves for SED modeling. These include 1.4 solar
metallicity ($Z_{\odot}=0.014$) coupled with the \citet{calzetti2000} attenuation curve (hereafter ``1.4 $Z_{\odot}+$Calzetti"), and 0.27 solar models accompanied by
the SMC extinction curve of \citet{gordon2003} (hereafter ``0.27 $Z_{\odot}$+SMC"). We also assumed
star-formation histories of the form $SFR(t)\propto t \times exp(-t/\tau)$ (i.e., delayed-$\tau$ models).
Here, $t$ is time since the onset of star formation and $\tau$ is
the characteristic star-formation timescale. The choice of
1.4 $Z_{\odot}+$Calzetti or 0.27 $Z_{\odot}$+SMC was assigned to each galaxy
based on the combination that yielded the lower $\chi^2$. Accordingly, 76 galaxies were modeled assuming 1.4 $Z_{\odot}+$Calzetti, and 18 galaxies with 0.27 $Z_{\odot}$+SMC. One filler target in GOODS-N at $z=1.81$ does not have a robust SED.

Finally, all relevant
photometric bands were corrected for the contributions from strong nebular
emission lines using the method described in \citet{sanders2021}. Photometry was additionally corrected for contributions from nebular continuum emission based on predictions from grids of Cloudy photoionization models \citep{ferland2017}, tied to the measured H$\beta$ line flux \citep[see][for full details]{sanders2024b}. 

We compared the stellar masses inferred from the SED modelling described above to those obtained by fitting the uncorrected broadband photometry using \textsc{prospector} \citep{johnson2021}, assuming a non-parametric star-formation history (SFH), and incorporating nebular emission in the SED models (see Topping et al. 2024, in prep., for a full description). Briefly, for this model setup, we assumed a SFH comprising eight independent time bins constrained by a continuity prior \citep[see, e.g.,][]{tacchella2022}. In addition, we fixed the nebular and stellar metallicities to be identical, and allowed the ionization parameter to vary from $\log(U) = -4$ to $-1$, with a uniform prior.
The stellar masses inferred from these non-parametric fits that consider nebular emission are in good agreement with those from the stellar-only FAST models, such that \textsc{prospector} yielded 0.09 dex larger masses on average, with a scatter between the masses from two methods of 0.45 dex for the full sample.

To place the AURORA sample in context, we plot the SFRs and stellar masses of our targets in Figure~\ref{fig:ms-ha}.  Stellar masses are derived as described above. In this figure, SFR is estimated from dust-corrected Balmer-emission-line fluxes. For the majority of galaxies, the H$\alpha$/H$\beta$ ratio was used to calculate $E(B-V)_{\rm gas}$, assuming the \citet{cardelli1989} dust extinction curve and an intrinsic ratio
of H$\alpha$/H$\beta=2.79$. In a small number of cases where
H$\alpha$ or H$\beta$ was not measured, we used, respectively, the ratio  H$\beta$/H$\gamma$ (assuming an intrinsic ratio of H$\beta$/H$\gamma=2.11$) or H$\alpha$/H$\gamma$ (assuming an intrinsic ratio of H$\alpha$/H$\gamma=5.90$) to estimate  $E(B-V)_{\rm gas}$. The value of $E(B-V)_{\rm gas}$ was used to infer the dust-corrected H$\alpha$ luminosity, and the conversion from \citet{hao2011}, adjusted to the \citeauthor{chabrier2003} IMF (${\rm SFR} (M_{\odot} {\rm yr}^{-1})= 10^{-41.33} L({\rm H}\alpha) ({\rm erg }{\rm s}^{-1})$), was used to translate H$\alpha$ luminosity to instantaneous SFR. Although the \citet{cardelli1989} curve was determined from observations within the Milky Way galaxy and the current work focuses on distant galaxies, we note the strong similarity between the \citet{cardelli1989} dust attenuation curve and the nebular attenuation curve determined for $z\sim 2$ star-forming galaxies by \citet{reddy2020} in the wavelength region where they overlap ($4000-6600$ \AA). Given that the $z\sim 2$ curve is limited to this overlapping wavelength region, while the \citet{cardelli1989} curve extends to both shorter and longer wavelengths where some of the key emission lines of interest fall, we adopt the \citet{cardelli1989} curve for the current work.

\subsection{AURORA Emission-line Sample}
\label{sec:obs-aurorasamp}
We investigate  several emission-line-ratio diagrams in this work. Each diagram is based on features spanning a different range of rest-frame wavelength. Given the wide redshift range of our targets and the wavelength gaps that are present in the spectra of some AURORA galaxies due to the manner in which the spectra fell across the two NIRSpec detectors, the subset of galaxies plotted in each diagram with coverage of all required lines differs slightly.  
Furthermore, as the focus of this work is the properties of star-forming galaxies, we separately identify both AGNs and quiescent galaxies within the AURORA sample. As in previous work, AGNs are identified spectroscopically on the basis of broad Balmer-line emission and/or [NII]/H$\alpha >0.5$ \citep{shapley2023a} (5 galaxies). Quiescent galaxies are identified either based on their photometric SEDs (4 galaxies) or else spectroscopically on the basis of strong stellar absorption lines and a lack of strong emission lines arising from star formation (1 galaxy). Two objects are identified as both AGNs and quiescent. Accordingly, 87 AURORA galaxies remain in our star-forming sample, and appear in subsequent plots.

\begin{figure*}
\centering
\includegraphics[width=1.0\linewidth]{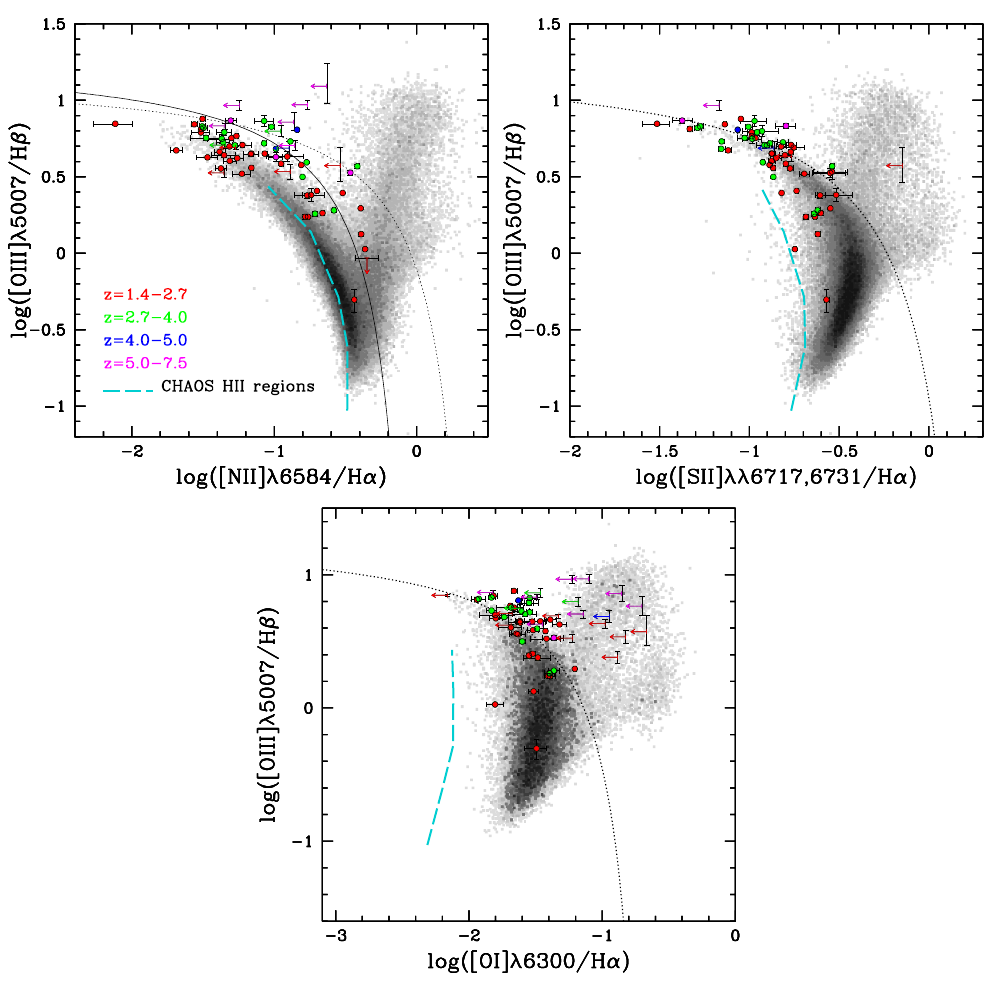}
\caption{``Classical" Emission-line Diagnostic Diagrams. In each panel, galaxies from the AURORA sample are plotted with colored symbols. Red, green, blue, and magenta are used, respectively, for the $1.4 \leq z < 2.7$, $2.7\leq z < 4.0$, $4.0 \leq z < 5.0$, and $z\geq 5.0$ samples. Star-forming galaxies with $\geq3\sigma$ detections of all 4 BPT emission lines are plotted with circles.  
Limits (3$\sigma$) are plotted as arrows. The 2D greyscale histogram and cyan dashed curve correspond, respectively, to the distribution of $z\sim 0$ SDSS galaxies and CHAOS H~II regions. {\bf Top left:} [NII] BPT diagram. The black dotted
curve is the ``maximum starburst" line from \citet{kewley2001}, while the black solid curve is an empirical
AGN/star-formation threshold from \citet{kauffmann2003}. {\bf Top Right:} [SII] BPT diagram. The black dotted
curve is the ``maximum starburst" line from \citet{kewley2001}. {\bf Bottom:} [OI] BPT diagram. The black dotted
curve is the ``maximum starburst" line from \citet{kewley2001}.  
}
\label{fig:bpt-panels}
\end{figure*}

\subsection{$z\sim 0$ Comparison Samples}
\label{sec:obs-z0samp}
As in past work, in order to understand the evolving emission-line ratios of distant galaxies, we selected comparison samples of both $z\sim 0$ galaxies and H~II regions. The $z\sim 0$ galaxy sample is drawn from the Sloan
Digital Sky Survey (SDSS) Data release 7
\citep[DR7;][]{abazajian2009}. Stellar-absorption-corrected emission-line measurements
are drawn from the MPA-JHU catalog of measurements
for DR7\footnote{Available at http://www.mpa-garching.mpg.de/SDSS/DR7/}. We limited the SDSS
sample to galaxies at $0.04\leq z\leq0.10$ to reduce aperture effects. SDSS does not have coverage of the [SIII]$\lambda\lambda9069,9532$ doublet, which we measure for AURORA galaxies at $z\leq 4.5$. Therefore, when investigating the relationship between [OIII]$\lambda 5007$ and [SIII]/[SII], we also fold in data from the Mapping Nearby Galaxies at APO
(MaNGA) integral field spectroscopic survey \citep{bundy2015}, which extends out to $1.04$~$\mu$m. We include the properties of 1150  MaNGA targets at $z<0.08$, such that [SIII] is covered, and sum a subset of pixels in the central 3\arcsec  for each galaxy to simulate a SDSS fiber spectrum \citep[see][for details]{sanders2020a}. 

For some of the emission-line ratio diagrams presented here -- in particular, those that include [OI] and [SII] emission -- galaxies and H~II regions occupy different regions of parameter space \citep{zhang2017}. Such differences arise because the SDSS emission-line spectra of $z\sim 0$ galaxies include the contributions of both H~II regions and diffuse ionized gas (DIG). As demonstrated in \citet{sanders2017} and \citet{shapley2019}, due to their $\sim 2$ orders of magnitude higher surface density of star formation ($\Sigma_{{\rm SFR}}$), and the anti-correlation between $\Sigma_{{\rm SFR}}$ and DIG fractional emission-line contribution to the integrated spectrum \citep{oey2007}, the integrated spectra of $z>2$ galaxies likely contain minimal contributions from DIG. Therefore, it is essential to compare the emission-line ratios of $z>2$ galaxies with those of $z\sim 0$ H~II regions. For this comparison, we assemble measurements for a sample of 298 H~II regions from the CHemical Abundances Of Spirals (CHAOS) survey \citep{berg2015,berg2020,croxall2015,croxall2016,rogers2021}.

\section{Results}
\label{sec:results}
{\it JWST}/NIRSpec boasts a 
transformative advance in both sensitivity and wavelength coverage. These developments facilitate new insights into ``classical" emission-line diagrams such as [OIII]$\lambda5007$/H$\beta$ vs. [NII]$\lambda6583$/H$\alpha$, [OIII]$\lambda5007$/H$\beta$ vs. [SII]$\lambda\lambda6717,6731$/H$\alpha$, [OIII]$\lambda5007$/H$\beta$ vs.
[OI]$\lambda6300$/H$\alpha$, and [OIII]$\lambda5007$/[OII]$\lambda3727$ (O$_{32}$) vs. ([OII]$\lambda3727$+[OIII]$\lambda\lambda4959,5007$)/H$\beta$ (R$_{23}$). Notably, for the first time we construct these diagrams including $\geq 3\sigma$ detections for {\it the vast majority} (i.e., $\geq 80$\%) of the individual galaxies in the AURORA sample with coverage of the relevant lines. In past work \citep[e.g.,][]{shapley2015,strom2017,cameron2023}, the corresponding all-line detection fraction was significantly lower. For example, in the CEERS NIRSpec analysis of \citet{sanders2023b}, it was  $\leq 50$\% in diagrams involving [NII] or [SII], and $\leq 15$\% when [OI] was being considered.   
In the face of such low detection fractions, composite spectra were used to obtain an unbiased estimate of the average sample properties. Because of the high detection fraction in AURORA, there is no need to rely on composite spectra.

In addition, it is possible to explore new parameter spaces, including emission lines at bluer rest-frame optical wavelengths that are useful for characterizing the highest-redshift galaxies \citep[e.g.,][]{bunker2023}. Such features (e.g., [NeIII]$\lambda3869$, H$\delta$) did not typically yield sufficient S/N for analysis in high-redshift galaxy spectra, pre-{\it JWST}. Finally, the extension of wavelength coverage out to 5$\mu$m enables a systematic investigation of emission lines at redder rest-frame wavelengths for galaxies at $z\leq 3$, such as [SIII]$\lambda\lambda9069,9532$ \citep{sanders2020a}, He~I$\lambda 1.083 \mu$m, [FeII]$\lambda 1.257\mu$m, and the Paschen series \citep{brinchmann2023,calabro2023}. We begin this section by revisiting ``classical" emission-line diagrams, and then turn to more novel parameter spaces at both shorter and longer rest-frame wavelengths. For closely-spaced emission-line ratios ([OIII]$\lambda5007$/H$\beta$, [NII]$\lambda6583$/H$\alpha$, [SII]$\lambda\lambda6717,6731$/H$\alpha$, 
[OI]$\lambda6300$/H$\alpha$) and those in the rest-frame near-IR, we do not apply a dust correction. However, for $O_{32}$, $R_{23}$, [NeIII]$\lambda 3869$/[OII]$\lambda3727$, ([NeIII]$\lambda3869$+[OII]$\lambda3727$)/H$\delta$, and $S_{32}$, we use dust-corrected line ratios based on $E(B-V)_{\rm gas}$, inferred as described in Section~\ref{sec:obs-sed}.

\subsection{``Classical" Emission-line Diagrams}
\label{sec:results-classical}

\subsubsection{BPT Diagrams}
\label{sec:results-classical-bpt}
We begin by revisiting familiar emission-line diagnostic diagrams, introduced by \citet{baldwin1981} (hereafter BPT) and \citet{veilleux1987} to classify extragalactic emission-line spectra as gas photoionized by young or evolved stars, shocks, or an AGN. Figure~\ref{fig:bpt-panels} includes plots of [OIII]$\lambda5007$/H$\beta$ vs. [NII]$\lambda6583$/H$\alpha$ (hereafter, the ``[NII] BPT diagram", top left), [OIII]$\lambda5007$/H$\beta$ vs. [SII]$\lambda\lambda6717,6731$/H$\alpha$ (hereafter, the ``[SII] BPT diagram", top right), and [OIII]$\lambda5007$/H$\beta$ vs.
[OI]$\lambda6300$/H$\alpha$ (hereafter, the ``[OI] BPT diagram", bottom). In each panel, the $z\sim 0$ SDSS sample is plotted as a 2D greyscale histogram in the background, and the median sequence of nearby extragalactic H~II regions from the CHAOS survey \citep[e.g.,][]{berg2015} is indicated as a dashed cyan curve. In the AURORA sample, we plot both detections and ``useful" limits (i.e., in which at least one line ratio has detections in both numerator and denominator).

The AURORA sample recovers the well-known offset in the [NII] BPT diagram for $z>1$ galaxies toward higher [NII]$\lambda6583$/H$\alpha$ and [OIII]$\lambda5007$/H$\beta$ relative to the sequence of $z\sim 0$ star-forming galaxies \citep{steidel2014,shapley2015}.
With smaller error bars than in previous work, the sequence of AURORA datapoints clearly scatters around a locus that is disjoint from both the highest density region of $z\sim 0$ the SDSS sequence, and the median of local HII regions. 
Prior to {\it JWST}, the long-wavelength cut-off at $\sim 2.4 \mu{\rm m}$ for ground-based near-IR spectroscopy limited studies of this diagram to galaxies at $z<2.6$. Early work based on composite {\it JWST}/NIRSpec spectra suggested a similar, constant offset at $z>2$ \citep{sanders2023b}, yet this result was based on a single stacked datapoint at each of $z=2.0-2.7$, $z=2.7-4.0$, $z=4.0-5.0$, and $z=5.0-6.5$. The AURORA sample includes 35 (17) star-forming galaxies at $z=1.4-2.7$ ($z=2.7-4.0$) with individual [NII] BPT measurements (as well as a small additional sample at $z>4$), enabling the first robust evolutionary measurement beyond $z\sim 2$. In contrast to the initial stacked result from \citet{sanders2023b}, we find that the $z=2.7-4.0$ sample (green points) is even further offset on average (by $\sim 0.1$~dex higher in typical [OIII]$\lambda5007$/H$\beta$ at fixed [NII]$\lambda6583$/H$\alpha$)  from the $z\sim 0$
star-forming sequence than the $z=1.4-2.7$ sample (red points). As shown in Figure~\ref{fig:ms-ha}, the $z=1.4-2.7$ and $z=2.7-4.0$ samples significantly overlap in terms of SFR and $M_*$. Therefore, this relative offset in the [NII] BPT diagram  suggests specific differences in the physical properties of H~II regions as opposed to global sample differences, but larger sample numbers are still required to overcome potential sample variance. 

\begin{figure}
\centering
\includegraphics[width=1.0\linewidth]{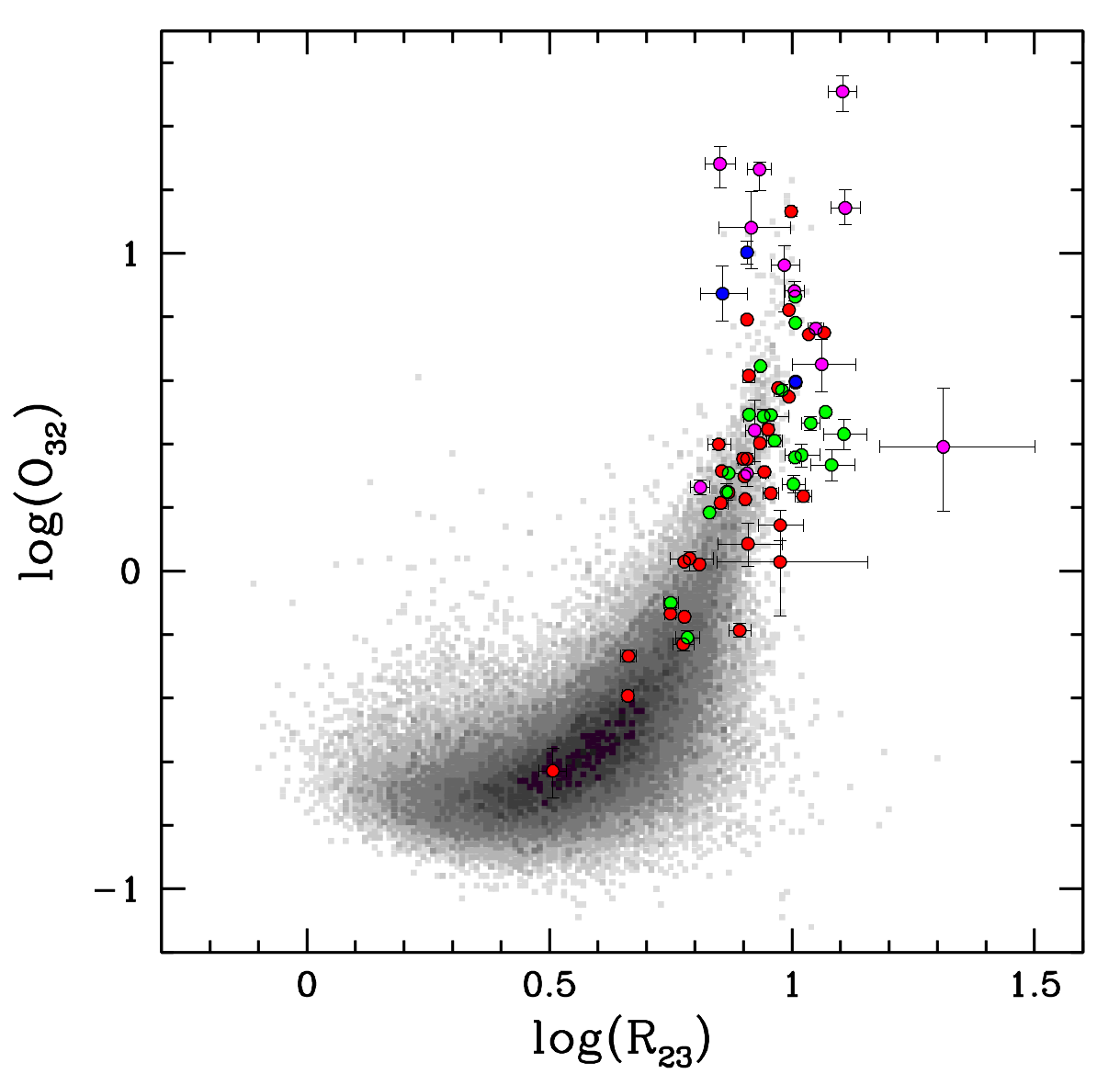}
\caption{$O_{32}$ vs. $R_{23}$ diagram, corrected for dust. Symbols for AURORA galaxies are as in Figure~\ref{fig:bpt-panels}. The $z\sim 0$ SDSS histogram only includes galaxies classified as star-forming using the criteria of \citet{kauffmann2003}, given that AGNs do not neatly segregate in this diagram.
}
\label{fig:o32r23}
\end{figure}

The sample of AURORA measurements at $z>4$  (including 4 galaxies with all lines detected, and 7 with upper limits in [NII]$\lambda6583$/H$\alpha$) is too small to draw firm conclusions regarding continued evolution in the [NII] BPT diagram. Early work with {\it JWST}/NIRSpec suggests that the typical
[NII]$\lambda6583$/H$\alpha$ line ratios for galaxies with stellar masses ranging from $10^8-10^{10} M_{\odot}$ are quite modest at $z>4$, ranging from 0.03 to 0.06 \citep{shapley2023b,cameron2023}. However, within AURORA, we observe a large dynamic range in [NII]$\lambda6583$/H$\alpha$, even at $z>6$, where we find a galaxy at $z=6.73$ with [NII]$\lambda6583$/H$\alpha=0.3$ \citep{shapley2024}. We now require much larger samples in the [NII] BPT diagram at $z>4$, with similar S/N to that obtained in AURORA.

Next we turn to the [SII] BPT diagram.
As discussed in \citet{sanders2017} and \citet{shapley2019}, $z\sim 0$ SDSS galaxies and H~II regions occupy different excitation sequences in this space, because of the added contribution of diffuse ionized gas to SDSS integrated spectra. This separation between the histogram of SDSS galaxies and the median sequence of CHAOS H~II regions is clear in the top right panel of Figure~\ref{fig:bpt-panels}. 
As justified in Section~\ref{sec:obs-z0samp}, $z\sim 2$ galaxies should be compared with $z\sim 0$ H~II regions in the [SII] BPT diagram \citep{shapley2019}. Based on their even higher $\Sigma_{{\rm SFR}}$, $z>2$ galaxies should as well. Using the correct $z\sim 0$ comparison dataset (i.e., the median H~II region sequence), we find, as in \citet{shapley2019}, that $z>1.4$ galaxies are significantly offset towards higher [OIII]$\lambda5007$/H$\beta$ and [SII]$\lambda\lambda6717,6731$/H$\alpha$.

The bottom panel of Figure~\ref{fig:bpt-panels} shows the [OI] BPT diagram. Detections of [OI]$\lambda6300$ were only achieved recently for the first time for a small sample of star-forming galaxies at $z>1$ \citep{clarke2023,sanders2023a}, based on ultra-deep ground-based spectroscopy with Keck/MOSFIRE. Stacked JWST/NIRSpec spectra have been used to detect [OI]$\lambda6300$ at high redshift as well \citep{cameron2023,strom2023}. The AURORA sample contains individual [OI]$\lambda6300$ detections of  46 star-forming galaxies, 
representing both an order-of-magnitude improvement in sample size and a significant (factor of several) increase in the typical [OI]$\lambda6300$ detection S/N.  \citet{zhang2017} demonstrate that H~II regions and DIG are characterized by distinct sequences of [OI]$\lambda6300$/H$\alpha$ emission. In turn, we argue that high-redshift galaxies should be compared with $z\sim 0$ H~II regions in the [OI] BPT diagram, in analogy with our above approach to the [SII] BPT diagram. As in \citet{clarke2023}, the distribution of $z>1.4$ AURORA galaxies is significantly offset towards higher in the [OI]$\lambda6300$/H$\alpha$ relative to local H~II regions.

\subsubsection{Ionization-Metallicity Diagram}
\label{sec:results-classical-o32r23}

A common tool for investigating the combination of ionization parameter and metallicity in star-forming galaxies is the so-called $O_{32}$ vs. $R_{23}$ diagram (Figure~\ref{fig:o32r23}). As described previously, because of the wavelength separation between [OII]$\lambda3727$ and [OIII]$\lambda5007$, we estimate dust-corrected line ratios assuming the \citet{cardelli1989} dust curve. 
In the  $O_{32}$ vs. $R_{23}$ diagram, the $z\sim 0$ SDSS comparison sample only includes galaxies classified as star-forming according to the [OIII]$\lambda5007$/H$\beta$ vs. [NII]$\lambda6583$/H$\alpha$ diagram. We note that $z\sim 0$ galaxies and AGNs don't separate as cleanly in this space, with the AGNs identified in the $z\sim 0$ [NII] BPT diagram simply adding scatter roughly symmetrically around the star-forming sequence.

Because useful $O_{32}$ vs. $R_{23}$ measurements can be obtained with coverage up to only $\sim 5000$\AA, we are able to plot the majority ($>70$\%) of the $z>5$ population in the AURORA sample here, in contrast to the BPT diagrams of the previous subsection. The AURORA sample shows a clear progression towards higher average $O_{32}$, in bins of increasing redshift. Also, there are are 20 galaxies with 
$\log(R_{23})\geq 1$ ($28$\% of AURORA galaxies with all the relevant lines detected), sharing almost no local counterparts \citep{sanders2016,strom2017}. We discuss these high-$R_{23}$ galaxies further in Section~\ref{sec:discussion-alpha}.

\begin{figure}[t!]
\centering
\includegraphics[width=1.0\linewidth]{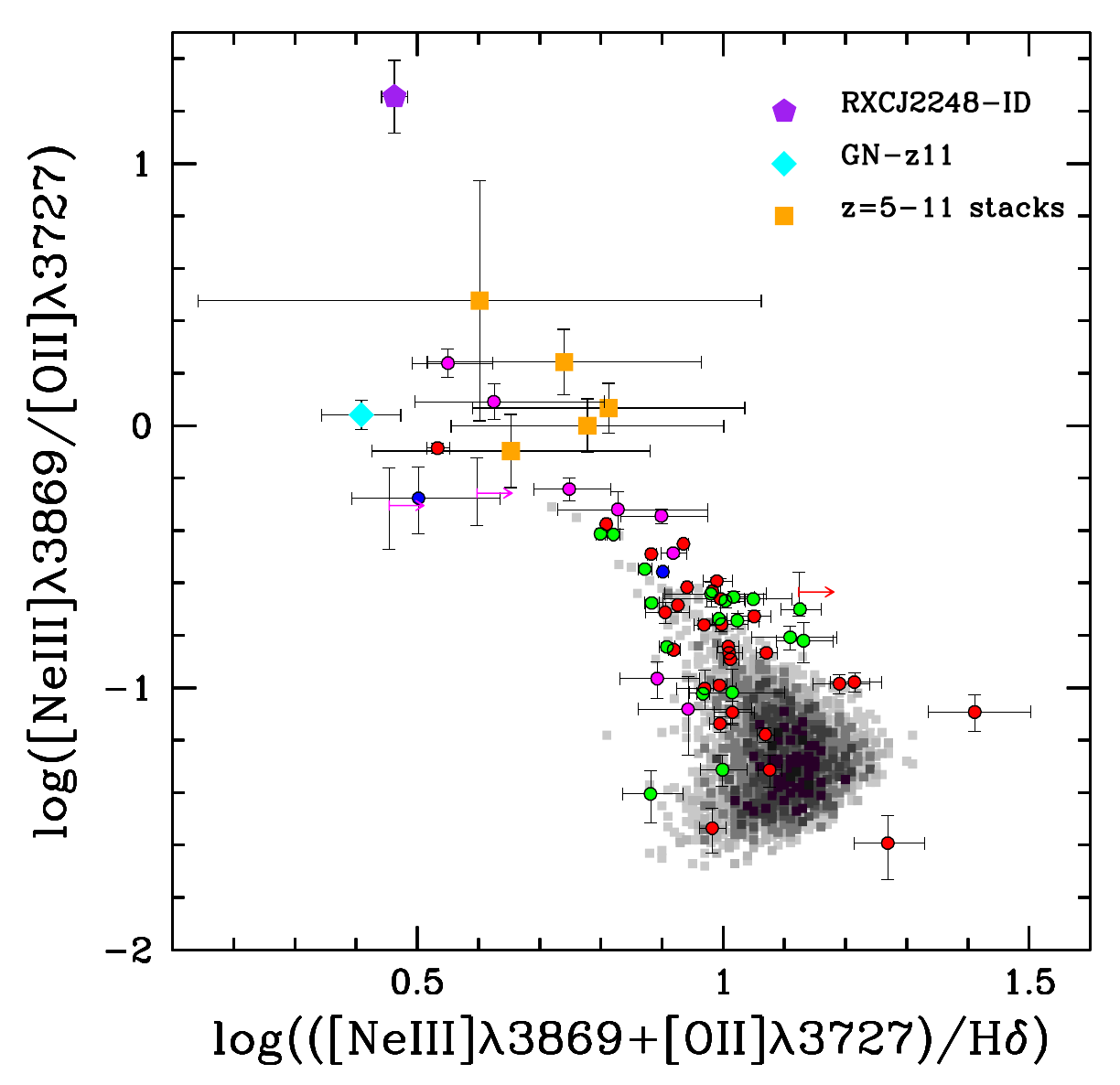}
\caption{[NeIII]$\lambda 3869$/[OII]$\lambda3727$ vs. ([NeIII]$\lambda3869$+[OII]$\lambda3727$)/H$\delta$, corrected for dust. Symbols for AURORA galaxies are as in Figure~\ref{fig:bpt-panels}. The $z\sim 0$ SDSS histogram only includes galaxies classified as star-forming using the criteria of \citet{kauffmann2003}, given that AGNs do not neatly segregate in this diagram. Also plotted are GN-z11 ($z=10.6$, cyan diamond) from \citet{bunker2023}, RXCJ2248-ID ($z=6.1$, purple pentagon) from \citet{topping2024}, and the stacked spectra at $z=5.5-9.5$ (orange squares) from \citet{robertsborsani2024}.
}
\label{fig:ne3o2n23}
\end{figure}

\subsection{A Bluer Ionization-Metallicity Diagram for $z>10$ Galaxies}
\label{sec:results-blue}

{\it JWST}/NIRSpec dramatically extended the redshift range over which diagnostic rest-optical lines can be observed. However, at $z>10$, where galaxies are now almost routinely being spectroscopically confirmed, even NIRSpec does not extend to strong, commonly-studied features such as [OIII]$\lambda5007$, H$\alpha$, and H$\beta$. Accordingly, in the context of understanding the physical properties of the ISM in the $z=10.6$ galaxy, GN-z11, \citet{bunker2023} introduce an emission-line diagnostic diagram based on bluer rest-frame optical features. In the joint space of ionization and metallicity, the [NeIII]$\lambda 3869$/[OII]$\lambda3727$ ratio serves the role of $O_{32}$, and, in place of $R_{23}$, there is ([NeIII]$\lambda3869$+[OII]$\lambda3727$)/H$\delta$. These emission lines are all close in wavelength, and their ratios therefore relatively insensitive to dust attenuation. \citeauthor{bunker2023} place GN-z11 in the context of photoionization models and local galaxies to infer a low metallicity and high ionization parameter, and indirectly compare with a small sample of $z>5$ galaxies with [NeIII]/[OII] and $R_{23}$ measurements (assuming a fixed ratio of H$\delta$/H$\beta$). GN-z11 appears to have similar line ratios to those of the stacked $z\sim 5.5-9.5$ NIRSpec prism (i.e., low-resolution) spectra of \citet{robertsborsani2024}. 
Thus far, however, there have been no direct comparisons between GN-z11 with individual galaxy spectra at high redshift and medium resolution.

In Figure~\ref{fig:ne3o2n23}, we plot [NeIII]$\lambda 3869$/[OII]$\lambda3727$ vs. ([NeIII]$\lambda3869$+[OII]$\lambda3727$)/H$\delta$ for the AURORA sample, along with the line ratios for GN-z11, the low-resolution stacks from \citet{robertsborsani2024}, and RXCJ2248-ID, a lensed $z=6.1$ galaxy with an extremely high S/N NIRSpec $R\sim1000$ spectrum \citep{topping2024}. Spanning from $z=1.4$ to $z=7.0$, at one extreme the AURORA star-forming galaxies overlap with the high [NeIII]$\lambda 3869$/[OII]$\lambda3727$, low ([NeIII]$\lambda3869$+[OII]$\lambda3727$)/H$\delta$ tail of the SDSS $z\sim 0$ distribution. At the other extreme, they extend towards GN-z11 and the $z\sim 5.5-9.5$ stacks. Within the AURORA sample, there exists a trend towards higher average 
[NeIII]$\lambda 3869$/[OII]$\lambda3727$ and lower average([NeIII]$\lambda3869$+[OII]$\lambda3727$)/H$\delta$ at increasing redshift. Accordingly, GN-z11 and the $z\sim 5.5-9.5$ stacks are most consistent with the $z>5$ portion of the AURORA sample. There are two exceptions at lower redshift: an auroral-line target, GOODSN-30274
($z=1.800$, red point), and a filler target GOODSN-917107 ($z=4.773$, blue point). These two galaxies have similar line ratios to those of GN-z11. Notably, the superior data quality of GOODSN-30274, including the detection of auroral [OIII]$\lambda4363$ emission, may provide insights into the detailed nature of GN-z11. 
RXCJ2248-ID has even more extreme [NeIII]$\lambda 3869$/[OII]$\lambda3727$ and lower average([NeIII]$\lambda3869$+[OII]$\lambda3727$)/H$\delta$ than GN-z11, with [NeIII]$\lambda 3869$/[OII]$\lambda3727$= $18.0\pm5.7$.
Understanding the extreme phase of galaxy formation represented by RXCJ2248-ID constitutes an important goal for studies of the galaxy population during the reionization epoch.

\subsection{Longer-Wavelength Diagnostics: [SIII] and into the Near-IR}
\label{sec:results-red}
In addition to enabling studies of rest-optical emission lines for galaxies at $z>10$, NIRSpec opens up a previously unexplored window into the rest-frame near-IR  spectroscopic properties of galaxies up to $z\sim 3$.

\subsubsection{$S_{32}$ Ratio}
\label{sec:results-red-s32}
\begin{figure}
\centering
\includegraphics[width=1.0\linewidth]{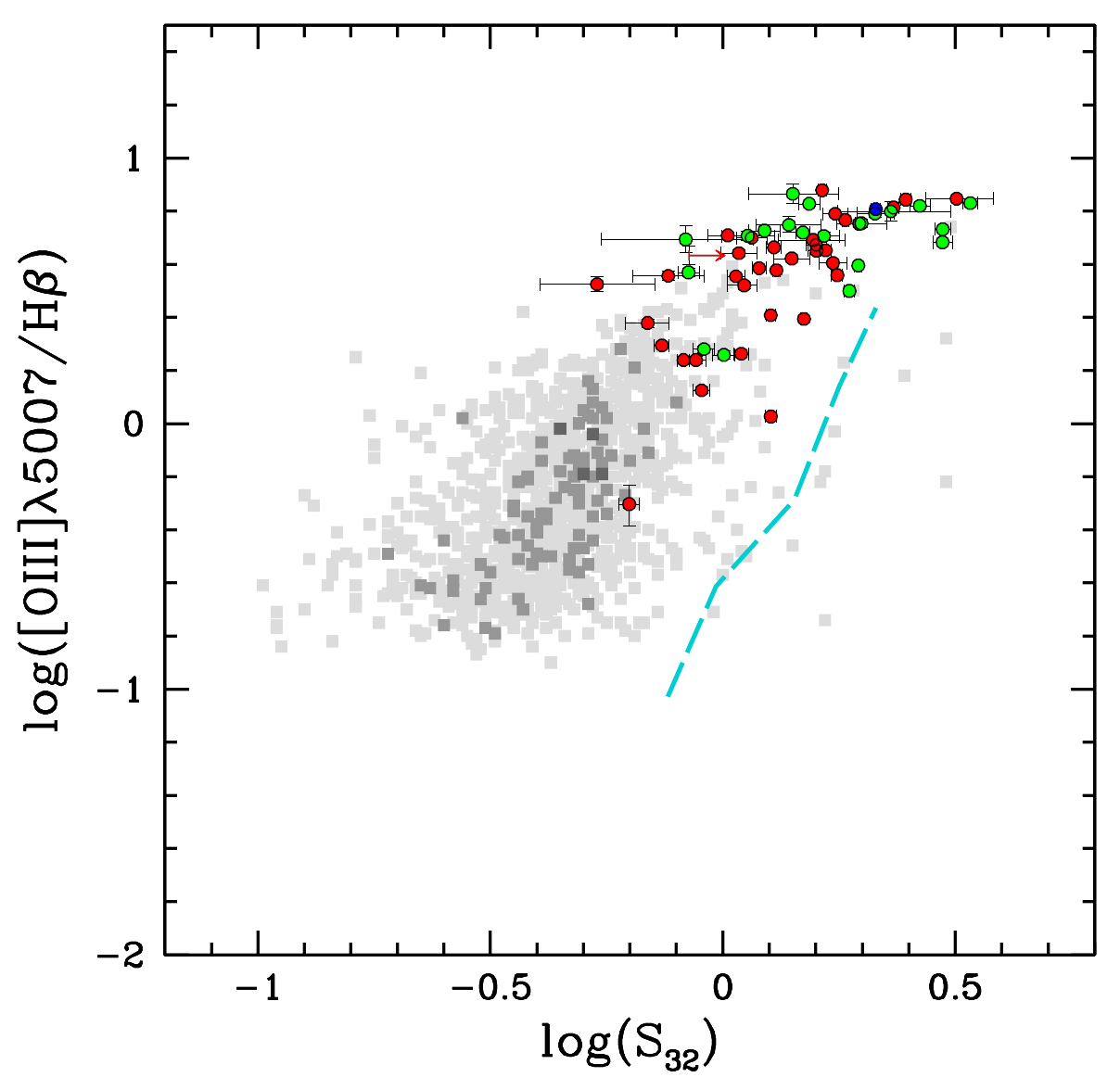}
\caption{[OIII]$\lambda5007$/H$\beta$ vs. $S_{32}$ (dust-corrected). Symbols for AURORA galaxies are as in Figure~\ref{fig:bpt-panels}.  The grey $z\sim 0$ 2D histogram is based on data from the MaNGA survey, which has coverage of [SIII]$\lambda\lambda9069,9532$, and only includes galaxies classified as star-forming using the criteria of \citet{kauffmann2003}. The cyan dashed curve represents the median sequence of H~II regions from the CHAOS survey.
}
\label{fig:o3s32}
\end{figure}

The ratio of [SIII]$\lambda\lambda9069,9532$/[SII]$\lambda\lambda 6717,6731$ ($S_{32}$) provides a complementary probe of the ionization parameter and has been used to probe the ionization state of the ISM in $z\sim 0$ galaxies \citep{kumari2021}.
As discussed in \citet{sanders2020a},  in concert with the [OIII]$\lambda 5007$/H$\beta$ ratio, $S_{32}$ can distinguish among different scenarios for the evolution of physical conditions in high-redshift star-forming regions. Specifically, it is possible to differentiate between the effects of an evolving ionizing spectrum or ionization parameter at fixed nebular metallicity, both of which have been suggested in the literature as possible explanations for the observed emission-line ratios in distant galaxies. In ground-based near-IR spectroscopic observations \citep[e.g.,][]{sanders2020a}, [SIII]$\lambda\lambda9069,9532$ was only detected in the K~band for a handful of individual $z\sim 1.5$ star-forming galaxies, and stacked composite spectra. In AURORA, we detect the $S_{32}$ ratio for 55 individual star-forming galaxies at $z=1.4-4.5$ (95\% of the targets with coverage of the relevant lines). In Figure~\ref{fig:o3s32}, we plot [OIII]$\lambda 5007$/H$\beta$ vs. the dust-corrected $S_{32}$ ratio for AURORA galaxies, along with galaxy-integrated measurements from the MaNGA survey \citep{bundy2015} and the median sequence of H~II regions from the CHAOS survey. As discussed previously, the CHAOS H~II regions comprise the appropriate DIG-free comparison sample for AURORA, in order to infer the redshift evolution of the internal conditions in H~II regions. The AURORA sample is unambiguously offset from the CHAOS sequence, towards lower $S_{32}$ at fixed  [OIII]$\lambda5007$/H$\beta$, within the range of overlapping [OIII]$\lambda5007$/H$\beta$.

\subsubsection{Rest-frame Near-IR Line Ratios}
\label{sec:results-red-nearir}

\begin{figure*}
\centering
\includegraphics[width=1.0\linewidth]{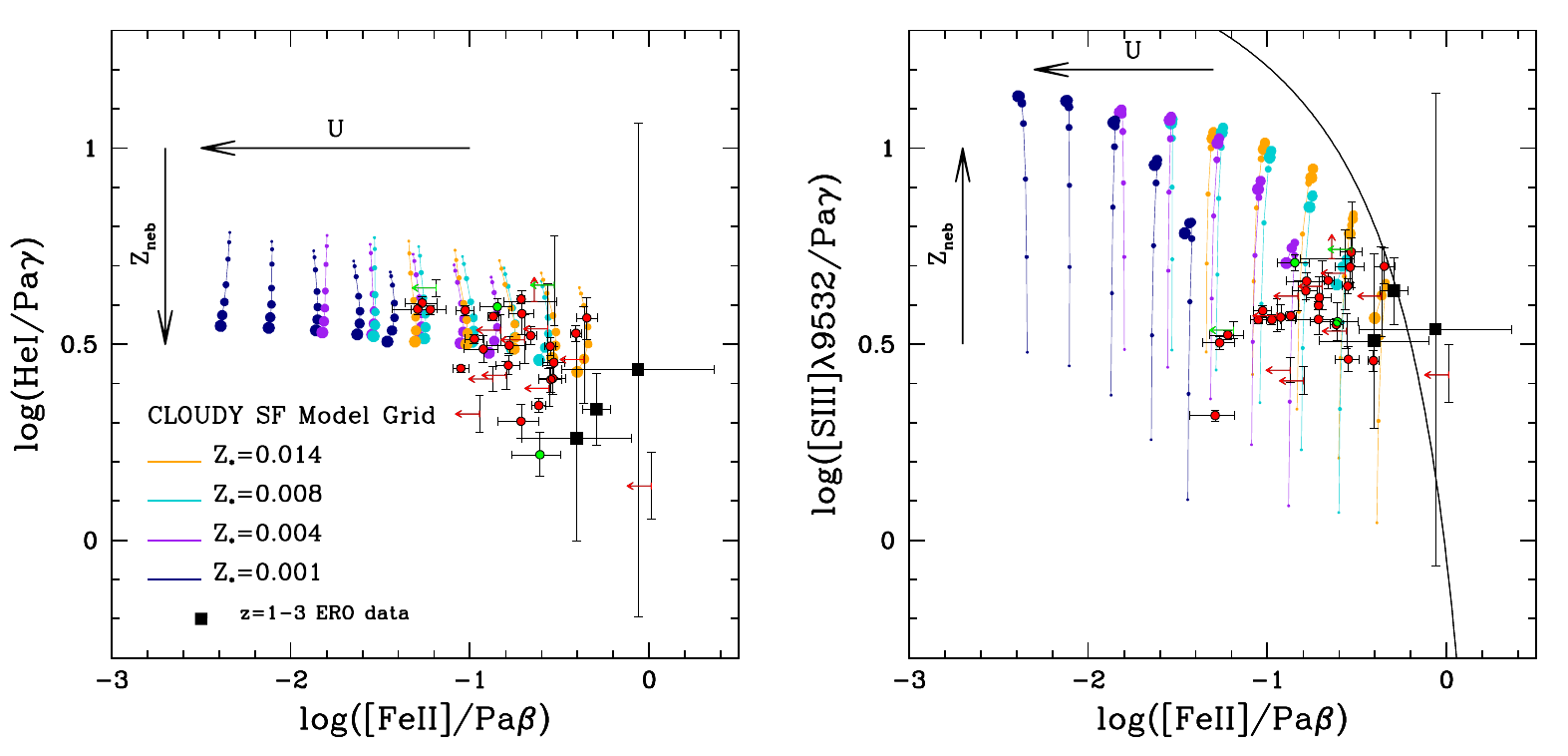}
\caption{Rest-frame Near-IR Emission-line Ratios. Symbols for AURORA galaxies are as in Figure~\ref{fig:bpt-panels}. Plotted in each panel are Cloudy \citep{ferland2017} model grids, assuming a BPASS \citep{eldridge2017} stellar population model with a constant star-formation history over $10^8$~yr for the ionizing spectrum. 
A dust depletion factor of 20 is assumed for Fe \citep{calabro2023,pettini2002}. 
Stellar metallicity (i.e., hardness of the ionizing spectrum) is indicated by the color of the model points, with $Z_*=0.001, 0.004, 0.008,$ and $0.014$ plotted, respectively, in dark blue, purple, turquoise, and orange. For connected model points of fixed ionization parameter ($U$) and stellar metallicity, increasing gas-phase metallicity ($Z_{\rm neb}$), ranging from 0.05 to 1.0 $Z_{\odot}$, is indicated with increasing point size. Arrows are used to indicate the direction in which  $U$ and $Z_{\rm neb}$ increase. The [FeII]$\lambda1.257\mu$m line flux is also scaled by the factor $Z_*/Z_{\rm neb}$ to reflect the $\alpha$-enhanced chemical abundance patterns in distant star-forming galaxies. 
{\bf Left:} He~I$\lambda1.083\mu$m/Pa$\gamma$ vs. [FeII]$\lambda1.257\mu$m/Pa$\beta$. In addition to the AURORA points and Cloudy model grids, we indicate three $z\sim 1-3$ galaxies from \citet{brinchmann2023} using black squares. {\bf Right:} [SIII]$\lambda9532$/Pa$\gamma$ vs. [FeII]$\lambda1.257\mu$m/Pa$\beta$. The same points are plotted as in the left panel, as well as the curve from \citet{calabro2023} intended to discriminate between star-forming galaxies
and AGNs.
}
\label{fig:ir-panels}
\end{figure*}

Both \citet{brinchmann2023}  and \citet{calabro2023} emphasized the potential of {\it JWST}/NIRSpec for detecting even longer rest-frame wavelength features with diagnostic power, including [SIII]$\lambda9532$, He~I$\lambda1.083$, Pa$\gamma$,  [FeII]$\lambda 1.257\mu$m, and Pa$\beta$. The ratios of such features have previously been used to determine the nature of the ionizing source in local galactic nuclei \citep[e.g.,][]{larkin1998,mouri2000,riffel2013}. In addition, [FeII]$\lambda 1.257\mu$m emission has been proposed as a tracer of the supernova rate in star-forming galaxies, due to its enhanced luminosity in the shocks associated with supernova remnants \citep[e.g.,][]{oliva1989,calzetti1997,rosenberg2012}.
Both of the above recent {\it JWST} works on rest-frame near-IR emission lines in distant galaxies feature the [FeII]$\lambda 1.257\mu$m/Pa$\beta$ ratio. \citet{brinchmann2023} pairs this ratio with He~I$\lambda1.083 \mu$m/Pa$\gamma$, and \citet{calabro2023}
combines it with [SIII]$\lambda9532$/Pa$\gamma$. The photoionization models featured in \citet{izotovthuan2016}, with an input stellar ionizing spectrum, suggest a typical maximum [FeII]$\lambda 1.257\mu$m/Pa$\beta=0.2$. Higher   [FeII]$\lambda 1.257\mu$m/Pa$\beta$ occurs with photoionization by the harder ionizing spectrum of an AGN, or else by fast interstellar shocks \citep{calabro2023,brinchmann2023}. In the [SIII]$\lambda9532$/Pa$\gamma$ vs. [FeII]$\lambda 1.257\mu$m/Pa$\beta$ diagram, \citet{calabro2023} determine the functional form for a curve discriminating between star-forming galaxies and AGNs (analogous to the curves in rest-optical BPT diagrams; e.g., \citealt{kauffmann2003}).

In Figure~\ref{fig:ir-panels}, we plot both 
He~I$\lambda1.083\mu$m/Pa$\gamma$ (left) and [SIII]$\lambda9532$/Pa$\gamma$ (right) vs. [FeII]$\lambda 1.257\mu$m/Pa$\beta$ for AURORA galaxies at $z\leq 3$. Local samples of such measurements are small \citep[e.g.,][]{calzetti1997,izotovthuan2016,riffel2019}, so we pair the AURORA measurements with Cloudy photoionization models \citep{ferland2017}, generated assuming a BPASS binary stellar population with constant star formation over $10^8$~yr for the ionizing spectrum \citep{eldridge2017}. Model ionization parameter runs from $\log(U)=-3.6$ to $-2.4$; model gas-phase metallicity extends from $Z_{\rm neb}=0.05-1.0 \: Z_{\odot}$; and model stellar metallicity ranges from $Z_*=0.001 - 0.014$. Given that the Cloudy models were run assuming a solar abundance pattern, we scale the [FeII]$\lambda 1.257\mu$m flux by a factor $Z_*/Z_{\rm neb}$ to reflect the well-established $\alpha$-enhancement in the chemical abundance patterns of high-redshift star-forming galaxies. Furthermore, a dust depletion factor of 20 was assumed for Fe (consistent with the model assumptions in \citealt{calabro2023} and high-redshift measurements in \citealt{pettini2002}). A different assumed dust depletion factor would correspond to a horizontal shift in the model grid. The AURORA dataset of 23 star-forming galaxies with individual detections of all relevant lines represents an order of magnitude increase in the $z>1$ sample size for this diagram. The 3 systems presented in \citet{brinchmann2023} have significantly larger error bars, and the sample presented in \citet{calabro2023} is dominated by individual non-detections. Furthermore, while the 3 $z=1-3$ systems from \citet{brinchmann2023} were presented as evidence for shock excitation, we find that the AURORA sample can easily be explained by photoionization by massive stars using this set of Cloudy models \citep{ferland2017}. For example, the median [FeII]$\lambda 1.257\mu$m/Pa$\beta$ is 0.2, which corresponds to [FeII]$\lambda 1.257\mu$m/H$\beta=0.03$, well within the range of stellar photoionization model predictions from \citet{izotovthuan2016} and \citet{calabro2023}. 

We note that a significant systematic uncertainty in the comparison between measurements and models is associated with the assumed dust depletion factor for Fe. Independent constraints on the dust depletion of Fe and its dependence on metallicity are essential for robust model comparisons. More fundamentally, but related, there is a systematic uncertainty associated with the atomic data (i.e., collision strengths, transition probabilities, and energy levels) used to predict [FeII] emission intensities in the photoionization models. For the purposes of comparison with the work of \citet{calabro2023}, we used the 2017 release of Cloudy \citep{ferland2017} and assumed a dust depletion factor of 20 for Fe. This release of Cloudy incorporates the model of \citet{verner1999} to describe the Fe~II ion. The 2023 release of Cloudy \citep{chatzikos2023} is updated with the model of \citet{smyth2019} for Fe~II as the default. The predicted [FeII]$\lambda 1.257\mu$m/Pa$\beta$ ratios are more than an order of magnitude lower using the \citet{smyth2019} description of Fe~II, and would therefore suggest that the AURORA dataset is most consistent with negligible Fe depletion onto dust. At the same time, other Fe~II models included with the Cloudy 2023 release \citep[e.g.,][]{tayal2018} predict [FeII]$\lambda 1.257\mu$m/Pa$\beta$ ratios consistent with those in \citet{verner1999}, and are therefore suggestive of substantial dust depletion. The significant systematic uncertainty in predicted near-IR [FeII] line intensities must be resolved for a robust interpretation of such features.

Finally, in purely empirical terms, we note that the median [FeII]$\lambda 1.257\mu$m/Pa$\beta$ ratio is lower than that inferred by \citet{calzetti1997} for a sample of local starbursts (i.e., 0.34). Understanding the origin of this difference will be an important component of constraining the origin of [FeII] emission in distant star-forming galaxies.

\section{Discussion}
\label{sec:discussion}

\subsection{The {\it JWST}/NIRSpec Case for $\alpha$-Enhancement}
\label{sec:discussion-alpha}
A physical picture was beginning to emerge from a decade of ground-based near-IR spectroscopic studies, explaining the rest-optical nebular emission properties of star-forming galaxies observed during the epoch of peak star-formation in the universe. With {\it JWST}/NIRSpec, it is possible to elucidate this physical picture with significantly increased wavelength coverage (and in turn redshift range) and S/N (and in turn detections for representative individual, rather than stacked, measurements). As first proposed in \citet{steidel2016}, and expanded upon in, e.g., \citet{strom2017}, \citet{shapley2019}, \citet{sanders2020a}, \citet{jeong2020}, \citet{runco2021}, \citet{topping2020}, \citet{clarke2023}, and \citet{cullen2021}, star-forming galaxies at $z\sim 2-3$ have rest-optical emission line properties suggestive of photoionization by chemically-young (i.e., $\alpha$-enhanced) massive stars. Accordingly, these stars have O/Fe ratios roughly 2-5 times higher than the ratio representative of the solar abundance pattern. This abundance pattern in massive stars corresponds to a harder ionizing spectrum (lower stellar iron abundance) at fixed nebular metallicity (gas-phase oxygen abundance). 

The AURORA data presented here confirm this physical picture in stunning detail. As described in Section~\ref{sec:results-classical-bpt} and Figure~\ref{fig:bpt-panels}, the AURORA sample is not only offset with respect to the star-forming sequence in the [NII] BPT diagram, but also relative to the sequence of local H~II regions in the [SII] and [OI] BPT diagrams. In \citet{shapley2019} and \citet{clarke2023}, we investigated the [SII] and [OI] BPT diagrams, with significantly lower S/N, and, in the case of the [OI] diagram in particular, a significantly smaller sample of individual measurements. The depth of the AURORA spectra yields detections for the majority (indeed, almost all) of galaxies targeted, and therefore the samples of individual galaxies plotted are not only larger but also {\it representative}. These representative samples indicate shifts in the emission-line diagnostic diagrams that are consistent with photoionization by a harder ionizing spectrum at fixed nebular metallicity, relative to what is observed in low-redshift star-forming galaxies. 

Furthermore, the high S/N of individual AURORA measurements enables us to discern differences between the average [NII] BPT properties of $z\sim2$ and $z\sim 3$ galaxies for the first time. We find that our $z\sim 3$ subsample is even more offset on average from the $z\sim 0$ star-forming in the [NII] BPT diagram than our $z\sim 2$ sequence is -- a difference previously impossible to detect based on shallower, stacked measurements \citep{sanders2023b}. This ongoing evolution before $z\sim 2$ may provide important, complementary clues regarding the early star-formation histories of distant galaxies \citep{cullen2021,kriek2019}. Given the design of the AURORA program, its target redshift distribution is skewed towards $z=1.4-4$  (Figure~\ref{fig:zhist}). A clear extension of the current work is to assemble larger samples of {\it JWST}/NIRSpec spectra of $z\sim 4-6.5$ star-forming galaxies with the same wavelength coverage and at least comparable depth. Such data will enable a similarly in-depth investigation of the evolution in the [NII], [SII], and [OI] BPT diagrams back to cosmic dawn.

The $S_{32}$ ratio is less familiar in the literature, based on the longer rest-frame wavelength of the [SIII] emission lines. \citet{sanders2020a} demonstrated how the combination of [OIII]$\lambda 5007$/H$\beta$ and $S_{32}$ line ratios could be used to unambiguously determine the physical conditions in distant star-forming galaxies in relation to their local counterparts. Specifically, based on Cloudy photoionization modeling, \citet{sanders2020a} showed that a shift in the [OIII]$\lambda 5007$/H$\beta$ vs. $S_{32}$ perpendicular to the excitation sequence of $z\sim 0$ H~II regions (the appropriate comparison sample for $z>1.4$ galaxies) could only be explained in terms of a harder ionizing spectrum at fixed nebular metallicity -- not a larger ionization parameter \citep[e.g.,][]{hirschmann2017,kashino2017}. This earlier work was based on a sample of only 10 individual detections at $z\sim 1.5$ with modest S/N, as well as stacked spectra containing $\sim 10-30$ $z\sim 1.5$ galaxies. In contrast, the AURORA sample contains 55 individual $S_{32}$ detections at $z=1.4-4.5$ with significantly higher S/N -- and robustly recovers the offset towards lower $S_{32}$  at fixed [OIII]$\lambda 5007$/H$\beta$ indicative of a harder ionizing spectrum at fixed nebular metallicity.

Another result from ground-based near-IR spectroscopic studies of $z\sim 2-3$ star-forming galaxies is the measurement of $\log(R_{23})$ values in excess of unity \citep{sanders2016,strom2017,runco2021}. There are virtually no corresponding measurements of local star-forming galaxies at such high $\log(R_{23})$ values \citep[but see, e.g.,][]{izotov2021}.  However, typical error bars on individual ground-based datapoints still make them consistent with the local star-forming sequence in the $O_{32}$ vs. $R_{23}$ diagram. With AURORA, the error bars on individual datapoints are small enough to demonstrate that 15 out of 20 of the galaxies in our sample with $\log(R_{23})>1$ also have the lower 68th-percent confidence bound  on $\log(R_{23})$ greater than 1. Such high $\log(R_{23})$ values with few if any local counterparts must be understood if $R_{23}$ is to be used as a robust metallicity indicator at $z\geq 4$, as in early JWST/{\it NIRSpec} work \citep{nakajima2023}. Notably, hardening the ionizing spectrum at fixed nebular metallicity in photoionization models produces higher peak $R_{23}$ values at fixed $O_{32}$ in photoionization models \citep[e.g.][]{steidel2016,sanders2020b}. Detailed photoionization modeling of the ionization parameter and ionizing spectrum of individual AURORA galaxies with direct metallicity estimates, as in \citet{sanders2020b}, will enable further examination of this question.

\subsection{Dust Depletion of Refractory Elements}
\label{sec:discussion-dust}
As presented in Section~\ref{sec:results-red-nearir}, {\it JWST}/NIRSpec enables measurements of rest-frame near-IR diagnostic line ratios for AURORA galaxies at $z\leq 3$. The range of [FeII]$\lambda 1.257\mu$m/Pa$\beta$ line ratios in the AURORA sample appears can be explained in terms of photoionization by massive stars -- not shocks or an AGN \citep{brinchmann2023}. However, the main source of uncertainty in comparing these new rest-frame near-IR line ratios with photoionization models is the assumption of the degree of dust depletion for the refractory element, iron. We assumed a factor of 20 depletion for iron, roughly consistent with previous work \citep{calabro2023,pettini2002}. However, better observational constraints on the level of dust depletion for iron in distant star-forming galaxies is an essential component of interpreting these new line ratios, building on the work of \citet{pettini2002}, \citet{jones2018}, and \citet{decia2016}. In particular, a description of dust depletion incorporating its metallicity dependence \citep{ izotov2006,rodriguez2005,gunasekera2023,romanduval2022} would represent an advance over the simple description presented here and in other recent work \citep[e.g.,][]{calabro2023}.

\subsection{Why Emission-line Diagrams?}
\label{sec:discussion-diagrams}
NIRSpec provides a significantly improved data quality for individual galaxy spectra compared to previous instrumentation. NIRSpec spectra from the AURORA program include detections of faint auroral lines for roughly half the sample ($\sim 50$ galaxies) enabling direct metallicities (Sanders et al. 2024, in prep.). In some cases, these spectra feature a large suite of up to tens of nebular emission lines spanning from the rest-frame blue optical into the rest-frame near-IR (e.g., Figure~\ref{fig:spec-30564}), which can be simultaneously modeled to infer the physical properties of the ionized ISM. This transformational advance in S/N raises a question of the very relevance of the emission-line diagnostic diagrams forming the focus of this paper. However, while the AURORA data demonstrate what is {\it possible} to achieve for individual galaxy spectra with NIRSpec with 24-hour exposure times, the majority of NIRSpec spectra collected to date do not have the same depth. These shallower NIRSpec spectra \citep{bunker2023,deugenio2024,shapley2023a} do contain detections of subsets of the strongest rest-frame optical emission lines, and therefore can be situated in emission-line diagnostic diagrams. Accordingly, the AURORA dataset -- for which both detailed modeling of individual spectra {\it and} plots of emission-line diagnostic diagrams are possible -- provides a physical basis for interpreting the emission-line diagnostic diagrams of the much larger samples of lower-S/N {\it JWST}/NIRSpec spectra that have been and will be collected in the years to come. In the current work, we present the AURORA emission-line diagrams. In future work, we will present detailed modeling of the ISM physical properties implied by individual AURORA spectra. Combining these complementary approaches is essential for charting the evolution of galaxy chemical enrichment and ISM physical conditions over cosmic time.

\section*{Acknowledgements}
This work is based on observations made with the NASA/ESA/CSA James Webb Space Telescope. The data were
obtained from the Mikulski Archive for Space Telescopes at
the Space Telescope Science Institute, which is operated by the
Association of Universities for Research in Astronomy, Inc.,
under NASA contract NAS5-03127 for JWST.  The specific observations analyzed can be accessed via \dataset[DOI: 10.17909/hvne-7139]{https://archive.stsci.edu/doi/resolve/resolve.html?doi=10.17909/hvne-7139}.
We also acknowledge support from NASA grant JWST-GO-01914. FC acknowledges support from a UKRI Frontier Research Guarantee Grant (PI Cullen; grant reference: EP/X021025/1).
ACC thanks the Leverhulme Trust for their support via a Leverhulme Early Career Fellowship.  CTD, DJM, RJM, and JSD acknowledge the support of the Science and Technology Facilities Council. JSD also acknowledges the support of the Royal Society through a Royal Society Research Professorship.
RD acknowledges support from the Wolfson Research Merit Award program of the U.K. Royal Society.  
KG acknowledges support from the Australian Research Council Laureate Fellowship FL180100060.
MK acknowledges funding from the Dutch Research Council (NWO) through the award of the Vici grant VI.C.222.047 (project 2010007169).
PO acknowledges the Swiss State Secretariat for Education, Research and Innovation (SERI) under contract number MB22.00072, as well as from the Swiss National Science Foundation (SNSF) through project grant 200020$\_$207349.
AJP was generously supported by a Carnegie Fellowship through the Carnegie Observatories. We acknowledge useful insights from Claus Leitherer, and a constructive and helpful report from an anonymous referee.
Finally, we thank members of the JADES team for assistance with target selection in the GOODS-N field.

\bibliographystyle{apj}
\bibliography{aurora-lineratio}

\end{document}